\documentclass[aps,pre,reprint,onecolumn,showpacs,superscriptaddress,notitlepage,11pt]{revtex4-2}
\usepackage{epsfig}
\usepackage{natbib}
\usepackage{ulem}
\usepackage{xspace}
\usepackage{xcolor}
\usepackage{graphicx}
\usepackage{amssymb}
\usepackage{cancel}
\usepackage{ulem}
\usepackage{lineno}
\usepackage{amsmath}
\usepackage{stackrel}
\usepackage{appendix}
\usepackage{url}
 \usepackage{enumitem}
 \usepackage{hyperref}
 \usepackage{multirow}

\newcommand{\I}{\mathcal{I}}

\newcommand{\even}{\mathrm{even}}

\newcommand{\TP}{\text{TP}}

\begin{document}      
	
	% \title{A tale of three approaches: Stretched-exponential relaxation and its dynamical phase transition}
 \title{A tale of three approaches: dynamical phase transitions for weakly bound Brownian particles}

	\begin{abstract}
    We investigate a system of Brownian particles weakly bound by attractive parity-symmetric potentials that grow at large distances as $V(x) \sim |x|^\alpha$, with $0 < \alpha < 1$. The probability density function $P(x,t)$ 
    at long times reaches the Boltzmann-Gibbs equilibrium state, with all moments finite.  However, the system's relaxation is not exponential, as is usual for a confining system with a well-defined equilibrium, but instead follows a stretched exponential $e^{- \mathrm{const} \, t^\nu}$ with exponent $\nu=\alpha/(2+\alpha)$. This problem is studied from three perspectives. First, we propose a straightforward and general scaling rate-function solution for $P(x,t)$. This rate-function, which is an important tool from large deviation theory, also displays anomalous time scaling and a dynamical phase transition. Second, through the eigenfunctions of the Fokker-Planck operator, we obtain, using the WKB method, more complete solutions that reproduce the rate function approach. Finally, we show how the alternative path-integral formalism allows us to recover the same results, with the above rate-function being the solution of the classical Hamilton-Jacobi equation describing the most probable path. Properties such as parity, the role of initial conditions, and the dynamical phase transition are thoroughly studied in all three approaches.

	\end{abstract}
	
    \author{Lucianno Defaveri}
    \affiliation{Department of Physics, Bar-Ilan University, Ramat-Gan 52900, Israel.}

    % \ead{lacadef@gmail.com}

    \author{Eli Barkai}
    \affiliation{Department of Physics, Institute of Nanotechnology and Advanced Materials, Bar-Ilan University, Ramat-Gan 52900, Israel.}

    \author{David A. Kessler}
    \affiliation{Department of Physics, Bar-Ilan University, Ramat-Gan 52900, Israel.}

\maketitle

\section{Introduction}

    Brownian particles in contact with a heat bath at temperature $T$ that are also subject to a binding potential field $V(x)$ will at long times reach a stationary equilibrium state. In this equilibrium state, the particles attain the standard Boltzmann-Gibbs expression for the probability density function $P(x,t)$, that is,
    \begin{equation}
        P(x,t \to \infty ) = P_{\text{eq}}(x) = \frac{1}{Z}  e^{-\frac{V(x)}{k_B T}} \, , \label{eq:Peq}
    \end{equation}
    where $k_B T$ is the temperature times the Boltzmann constant and $Z$ is the Boltzmann-Gibbs normalizing partition function
    \begin{align}
        Z = \int_{-\infty}^{\infty} e^{-\frac{V(x)}{k_B T}} \, dx \, . \label{eq:Z-def}
    \end{align}
    In this work, we consider a class of even potentials behaving for large $x$ as $V(x) \approx V_0\, |x/\ell|^\alpha$. Here, $\alpha$ denotes the scaling exponent, $V_0$ is a positive constant indicating the strength of the potential, and $\ell$ represents a characteristic length scale. The time-dependent probability density of the aforementioned Brownian particles is described by the Fokker-Planck equation (FPE), 
    \begin{align}
        \frac{\partial}{\partial t} P(x,t) &= D \left\{  \frac{\partial^2}{\partial x^2} P(x,t) - \frac{\partial}{\partial x} \left[ \frac{F(x)}{k_B T} P(x,t) \right] \right\} \, , \label{eq:FPE}
    \end{align}
    where we have the force $F(x) = - V'(x)$ and $D$ is the diffusion coefficient.
    
    When $\alpha > 1$, the system exponentially relaxes towards the Boltzmann-Gibbs equilibrium state in Eq.\,(\ref{eq:Peq}). The rate of relaxation is governed by the first non-zero eigenvalue associated with the Fokker-Planck operator (see Eq.\,(\ref{eq:FPE-operator}) below). For such systems, the eigenvalue spectrum is discrete~\cite{Risken1996}, starting at 0, which is the time-independent equilibrium state, as in Eq.\,(\ref{eq:Peq}). We note for future reference that odd/even observables will, at long times, relax to their equilibrium values at different rates, governed respectively by the lowest nonzero odd/even eigenvalue.
    
    If $\alpha < 0$, the potential is not binding, the partition function in Eq.\,(\ref{eq:Z-def}) diverges and the equilibrium state is non-normalizable. Despite this, the Boltzmann-Gibbs factor $e^{-V(x)/k_B T}$ plays a key role in calculating averages, and such systems have been extensively studied using tools from infinite ergodic theory \cite{Aghion2019,Aghion2020}.
    
    The logarithmic potential limit is obtained when $\alpha \to 0$ (and $V_0=\tilde{V}_0/\alpha \to \infty$), and so $V(x) \sim \tilde{V}_0 \ln |x/\ell|$~\cite{Kessler2010,Dechant2011,Mukamel2011,Mukamel2012}. These systems possess a normalizable equilibrium state for  $\tilde{V}_0 / k_B T > 1$, with a power-law tail, leading to anomalous statistics and higher-order moments that grow as various powers of the time. This interesting case has also been studied extensively \cite{Barkai2023} due to its numerous applications to optical lattices and long-range systems.  For smaller $V_0$, on the other hand, the equilibrium state is non-normalizable and the behavior is similar to the $\alpha<0$ case.
    
\begin{figure}
            \centering
            \includegraphics[width=0.8\textwidth]{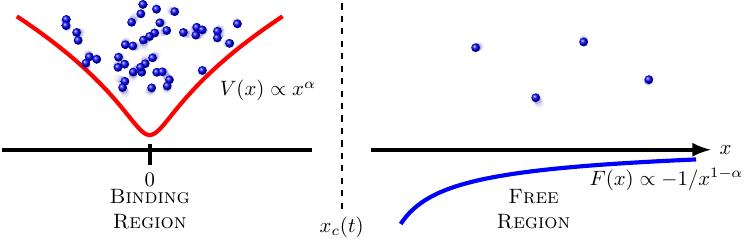}
            \caption{A schematic representation of the Brownian particles moving in space under the effect of a potential as in Eq.\,(\ref{eq:model-potential}). There is a binding region near the origin, where the effects of the potential are strongly felt (left of image). In this region, for a finite long time $t$, the probability will be approximately the equilibrium, as in Eq.\,(\ref{eq:Peq}). However, for $x>x_c(t)$, see Eq.\,(\ref{eq:def-zc}), the particles are in the free region, where the effects of the force are negligible, and their PDF has not converged to the equilibrium expression.}
            \label{fig:model-figure}
        \end{figure}

    In this paper, we consider the case of  $0 < \alpha < 1$. For concreteness, we will consider the family of potentials
	\begin{align}
	    V(x) &= V_0 \left( 1 + \frac{x^2}{\ell^2} \right)^{a/2} \, , \label{eq:model-potential}
	\end{align}
    though the results extend directly to the general case of even potentials which behave as $V(x)\sim x^\alpha (1 + {\cal O}(1/x))$ as $x\to \infty$. For this range of $\alpha$, we have that the potential is confining as $V(x\to \infty) \to \infty$. However, the force, which behaves as $F(x) = - \alpha (V_0/\ell) / (x/\ell)^{1-\alpha}$ decays to zero. This means that the particles will experience two distinct regions, a binding region for $x$ of order $\ell$, where the potential is strongly felt and a free region for large $x$, where the force is negligible and the motion approximates that of a freely diffusing particle. We present a schematic representation in Fig.\,\ref{fig:model-figure}.  For the sake of simplicity, throughout this paper, we will rescale position and time to dimensionless variables by using $x \to x/\ell$ and $t \to D t / \ell^2$.
    
    We shall see that for the family of potentials in Eq.\,(\ref{eq:model-potential}), the spectrum $\lambda_k$ of the relevant Fokker-Planck operator is mixed. As is the case of the logarithmic potential for $\tilde{V_0}/kT>1$, it has a continuous spectrum extending from zero along with a discrete normalized ground-state at eigenvalue zero. As opposed to the logarithmic case, however, for $0<\alpha<1$ the dominant contribution to $P(x,t)$ for large times comes not from around $\lambda_k=0$, but rather from the vicinity of a finite $\lambda_{k*}(t)$ which scales as a negative power of the time. This leads to a stretched-exponential relaxation to equilibrium \cite{Kohlrausch1854,Klafter1986,Kisslinger1993,Klafter2005,Mukherjee2023}. The existence of a continuum of modes has been long understood as a prerequisite for stretched-exponential behavior. However, it is not sufficient, as the logarithmic potential case demonstrates. The class of weak power-law potentials thus forms an interesting and natural example for stretched exponential relaxation. Interestingly, we will show that this relaxation is related to a dynamical phase transition. 
    
    Specific cases of this class of potentials have been studied to model the effective dynamics of the radius $r$ of spherical droplets in Ising ferromagnets bellow critical temperatures \cite{Fisher1987,Langer1989}. The effective dynamics of the radius follows a Langevin equation, with the effective potential scaling with the radius as $r^{2/3}$ ($\alpha = 2/3$) for $d=2$ dimensions, and $r$ ($\alpha=1$) for $d=3$ dimensions. The relaxation for $d=2$ display stretched exponential relaxation, with the same exponent as we will obtain in Sec. \ref{sec:moments}.

    This striking characteristic of the eigenvalue spectrum hints at the existence of an excited-mode scaling solution in the form of a rate function. In large deviation theory \cite{Touchette2009,Kafri2015,Kafri2018,Touchette2018,Meerson2019,Touchette2016,Touchette2017,Wang2020,Harris2020,Espigares2023a,Oren2023}, the rate function $\I$ is an important tool for studying the behavior of large fluctuations, and it is defined formally as
     \begin{align}
         \I(z) = - \lim\limits_{\stackrel{t,x\to\infty}{z=x/t^\gamma \textrm{fixed}}} \hspace{-0.1in} \frac{\ln P(x,t)}{t^\nu}  \, , \label{eq:rate-function-formal}
     \end{align}
    where $z \equiv x/t^\gamma$ is a scaled variable with $\nu$ and $\gamma$ being scaling exponents. For some systems, the rate function may contain nonanalytical points, marking the transition between two expressions for the rate function, and therefore, the statistics of the system. We call these points of dynamical phase transitions \cite{Garrahan2007,Garrahan2009,Majumdar2015,Smith2022b,Smith_cond_2022,Mori2022,Mori2023,Kanazawa2024}, due to the analogous relation between the rate function and the equilibrium free energy. We will explain the rate function approach in more detail in the next section.

    This manuscript is structured as follows: First, for completeness, in Sec. \ref{sec:scaling-solution} we present a rate-function scaling solution to the FPE\,(\ref{eq:FPE}). We will see that the rate function, which represents the leading exponential contribution of the PDF using the proper scaled variables is multi-valued. This multi-valued nature is explained through the parity and initial conditions. The physical meaning of the scaled variables is discussed in Sec. \ref{sec:scaled-vars}. In Sec. \ref{sec:eigenfunction-approach}, we use an eigenfunction expansion of the Fokker-Planck operator. This first-principles approach allows us to recover and extend all results gleaned through the scaling approach. In Sec.\,\ref{scaled-rate-eigen}, we show the equivalence between the scaling rate function and the eigenfunction solutions. In Sec. \ref{sec:V-shaped}, we analytically solve the particular case of the $V$-shaped potential, which is equivalent to the $\alpha=1$ case. This important model allows us to better understand the nature of the dynamical phase transition observed \cite{Garrahan2009, Touchette2016, Nyawo2018, Smith2022} from the perspective of an eigenfunction expansion. This is extended, in the following Sec. \ref{eq:phase-transition-general}, to the $0<\alpha<1$ range. In Sec. \ref{sec:moments} we use the PDF obtained through the eigenfunction approach to calculate the ensemble mean of observables. In Sec. \ref{sec:path-integral} we use the path integral solution of the FPE\,(\ref{eq:FPE}) to derive the dynamical phase transition. We also show how this method, which is widely used in the context of large deviation theory, can directly tackle parity problems. We finish with some concluding remarks. A brief summary of some of the results presented here appeared in \cite{Defaveri2024}.

    \section{Scaling solution \label{sec:scaling-solution}}

\begin{figure}
        \centering
        \includegraphics[width=0.9\textwidth]{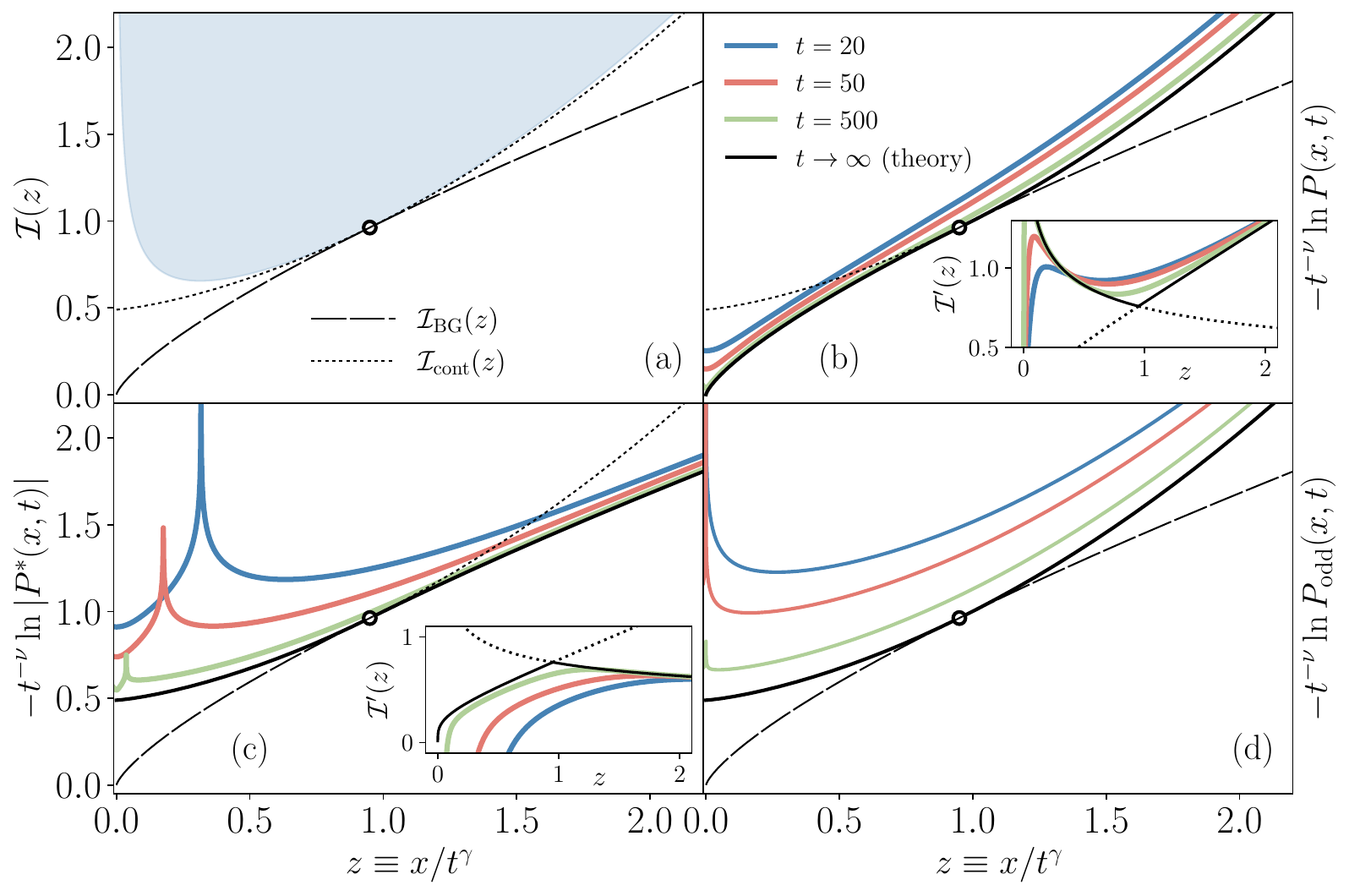}
        \caption{The numerical (finite-time) rate function (solid colored lines), compared to the theoretical predictions (dashed black lines for different times shown in the caption of the panel (b) (common to all panels). In panel (a) we have the two analytical rate functions $\I_\mathrm{BG}(z)$ and $\I_\mathrm{cont}(z)$ with the negative discriminant region highlighted. Panels (b), (c), and (d) represent the same rate functions as shown in the text (see beneath Eq.\,(\ref{eq:critical-point-via-rate-function})). On the inset of panels (b) and (c), we show the derivative of the solid lines in the main plot, with respect to $z$, where we can see the non-analytical nature of the dynamical phase transition clearly. This dynamical phase transition occurs when the system switches branches at the critical point, which we marked with a circle. For the equilibrium solution, panel (a), and for the odd part, panel (d), we do not have any critical point where a dynamical phase transition takes place. Hence, the existence or not of the phase transition depends on the initial condition and on the parity of the observable of interest. In all cases, we can clearly see that the numerical results approach the theoretical rate function for longer times.
        The solid theoretical lines are obtained by numerically integrating the FPE\,(\ref{eq:FPE}) with the potential in Eq.\,(\ref{eq:model-potential}) for $\beta=1$, $\alpha=3/4$ and $\ell = 1$. The peak observed in panel (c) is the point where the sign of $P^*$ changes from positive to negative.}
        \label{fig:rate-functions}
    \end{figure}

    In order to obtain a complete time-dependent solution of the FPE, Eq. \eqref{eq:FPE}, we must make use of methods such as the eigenfunction approach, which is a difficult task.  While the eigenfunction approach, which we will pursue in detail in the next section, is a viable method in the long-time limit, a scaling approach offers a simpler way to express the long-time PDF.
    For long times, we can define a new scaled position variable $z \equiv x/t^\gamma$ as we define in Eq.\,(\ref{eq:rate-function-formal}), and assume~\cite{Defaveri2024}
    \begin{align}
        P(x,t) \sim e^{-t^\nu \mathcal{I}(z)} \, , \label{eq:rate-function-Pxt}
    \end{align}
    where $\nu$ and $\gamma$ are exponents to be determined and $\mathcal{I}(z)$ plays the role of a rate function in the large-deviation formalism~\cite{Freidlin1998, Touchette2009}. However, since $\nu \neq 1$ and $\gamma \neq 1$, the rate function will be termed anomalous \cite{Touchette2016,Smith2022b}. By plugging Eq.\,(\ref{eq:rate-function-Pxt}) into the FPE (\ref{eq:FPE}) and considering only the leading terms in $1/t$, we obtain for the time derivative
    \begin{align}
        \frac{1}{P(x,t)} \frac{\partial}{\partial t} P(x,t) \to \frac{z \gamma \I'(z) - \nu \I(z)}{t^{1-\nu}} \, , \label{eq:time-FPE}
    \end{align}
    and for the Laplacian term
    \begin{align}
        \frac{1}{P(x,t)} \frac{\partial^2}{\partial x^2} P(x,t) \to \frac{\I'(z)^2}{t^{2 \gamma - 2 \nu}} \, , \label{eq:laplacian-FPE}
    \end{align}
    and finally, for the force field term
    \begin{align}
        \frac{1}{P(x,t)} \frac{\partial}{\partial x} \left[ \frac{F(x)}{k_B T}  P(x,t) \right] \to \frac{V_0}{k_B T}\frac{ \alpha }{z^{1-\alpha}} \frac{\I'(z)}{t^{(2-\alpha)\gamma - \nu}} \, . \label{eq:force-FPE}
    \end{align}
    We are interested in the scaling regime where all terms, and hence
    the force, the diffusion and the time-dependence contribute equally.  This rules out other possible choices whre ones of these effects is negligible. 
   For example, the trivial $\nu = \gamma = 0$ solution eliminates the time derivative term, and the rate function would assume the value $t^\nu \I(z) = V(x)/k_B T$, which is equivalent to the Boltzmann-Gibbs probability, see Eq.\,(\ref{eq:Peq}). Another example is $\nu = \gamma = 1$ where, for $\alpha < 1$, the force field term is smaller than the other two by $t^{1-\alpha}$, and the rate function becomes $\I(z) = z^2/4$. This describes diffusion in the absence of the force.
   Rather, for the exponents 
    \begin{align}
    \nu = \frac{\alpha}{2-\alpha} ~ \mathrm{and} ~ \gamma = \frac{1}{2-\alpha} \, . \label{eq:scaled-exponents}
    \end{align}
    we have that Eqs.\,(\ref{eq:time-FPE}), (\ref{eq:laplacian-FPE}) and (\ref{eq:force-FPE}) will have the same time dependence.
    This results in the non-linear odinary differential equation for $\I(z)$:
    \begin{align}
    \mathcal{I}'(z)^2 - \left( \frac{V_0}{k_B T} \frac{\alpha}{
    z^{1-\alpha}} + \frac{z}{2-\alpha} \right) \mathcal{I}'(z) + \frac{\alpha \mathcal{I}(z)}{2-\alpha} = 0 \, . \label{eq:df-rate-function}
    \end{align}
    Note that, because the previous equation is quadratic with $\I'(z)$, we have two possible differential equations to choose for every point. We may write them explicitly as
    \begin{align}
    \I'(z) &= \frac{1}{2}\left( \frac{V_0}{k_B T} \frac{\alpha}{
    z^{1-\alpha}} + \frac{z}{2-\alpha} \right) \pm \frac{1}{2} \sqrt{\left( \frac{V_0}{k_B T} \frac{ \alpha}{z^{1-\alpha}} + \frac{z}{2-\alpha} \right)^2 - \frac{4 \alpha \I(z)}{2 - \alpha}} \, . \label{eq:branches-Iprime}
    \end{align}
    In order for the solutions to be always real, we must impose that the discriminant $\Delta (z,\I)$ is always nonnegative, that is,
    \begin{align}
        \Delta (z,\I) = \left( \frac{V_0}{k_B T} \frac{ \alpha}{z^{1-\alpha}} + \frac{z}{2-\alpha} \right)^2 - \frac{4 \alpha \I}{2 - \alpha} \geq 0 \, ,
    \end{align}
    where we suppress the $z$ dependence, namely $\I = \I(z)$. The equality in the previous equation represents the case where we only have one viable solution. This region of negative discriminant is shown in Fig.\,\ref{fig:rate-functions}\,(a), for $\alpha=3/4$, with the boundary representing the zero-discriminant line.

    If ${\cal{I}}$ and ${\cal{I}}'$ are to be continuous, the only possibility to switch branches above is at a point where the discriminant $\Delta(z,{\cal{I}})$ vanishes, i.e. 
    \begin{align}
         \left( \frac{V_0}{k_B T} \frac{ \alpha}{z^{1-\alpha}} + \frac{z}{2-\alpha} \right)^2 - \frac{4 \alpha \I}{2 - \alpha} = 0 \, .
    \end{align}
    
    In general, there are many solutions to Eqs.\,(\ref{eq:branches-Iprime}). We now wish to find the solutions that obey certain restrictions: 1) $\I(z) \geq 0$ for all $z$, as negative values of $\I(z)$ would imply that the probability will grow exponentially as $t \to \infty$, and 2) $\Delta (z,\I) \geq 0$, as the solutions must be real, as mentioned. There are only two analytical solutions that respect both restrictions, one of which is equivalent to the Boltzmann-Gibbs ground state, that is $V(x)/k_B T \to t^\nu \mathcal{I}_\mathrm{BG}(z)$ and $\mathcal{I}_\mathrm{BG}(z) = V_0 z^{\alpha}/k_B T$. The second solution, which we will refer to as $\mathcal{I}_{\mathrm{cont}}(z)$, does not have a simple expression, but for large $z$ it grows asymptotically as it would for a free-particle, that is, $\I_\mathrm{cont}(z) \approx z^2/4$. We name this rate function $\I_\mathrm{cont}$ because, as we will see in detail through the eigenfunction approach, it is related to the contribution from the continuum excited modes. Further in this Section we will explain how this continuum rate function can be obtained through $P(x,t)$.

    Both solutions cross each other at a critical point $z_c$, which is also the point that both functions are tangent to the zero-discriminant line, see Fig.\,\ref{fig:regions}\,(a). At that point, both the actual functions $\mathcal{I}_\mathrm{BG}(z_c) = \mathcal{I}_\mathrm{cont}(z_c)$ and their first derivative are the same $\mathcal{I}_\mathrm{BG}'(z_c) = \mathcal{I}_\mathrm{cont}'(z_c)$. The critical point can be written explicitly as
    \begin{align}
        z_c = \left( \frac{\alpha(2-\alpha) V_0}{ k_B T} \right)^{1/(2-\alpha)} \label{eq:def-zc}  \, .
    \end{align}
    This allows us to write four continuous solutions, as at the critical point we may switch from one analytical expression to the other continuously. All four expressions have a physical meaning related to parity and initial conditions
    \begin{itemize}
        \item[(a)] Equilibrium distribution, $\I_\mathrm{eq}(z)$: if our system is placed at time $t=0$ in the Boltzmann-Gibbs equilibrium state, Eq\,(\ref{eq:Peq}), for all $x$, the only solution is $\mathcal{I}_\mathrm{BG}(z)$ for all $z$. Explicitly, the rate function for this PDF is written as
        \begin{align}
            \I_\mathrm{eq}(z) = - \lim\limits_{\stackrel{t,x\to\infty}{z=x/t^\gamma \textrm{fixed}}} \hspace{-0.1in}  \frac{\ln P_\mathrm{eq}(x,t)}{t^\nu} \, .
        \end{align}
        \item[(b)]  Localized initial condition, even part, $\I_\mathrm{even}(z)$: if our system is placed with localized initial condition, $P(x,0) = \delta(x-x_0)$,  then, the even part of the PDF, that is $P_\mathrm{even}(x,t) = (P(x,t) + P(-x,t))/2$ is described by $\mathcal{I}_\mathrm{BG}(z)$ for $z<z_c$. However, for $z>z_c$, the particles have not yet reached equilibrium, and the system behaves closely to a free-particle $\mathcal{I}_\mathrm{cont}(z)$, for $z>z_c$. At this point, we present this result from a mathematical point of view. However, in the next section we will provide a physical perspective to the transition. Explicitly, the rate function for this PDF is written as
        \begin{align}
            \I_\mathrm{even}(z) = - \lim\limits_{\stackrel{t,x\to\infty}{z=x/t^\gamma \textrm{fixed}}} \hspace{-0.1in}  \frac{\ln \left( \frac{P(x,t) + P(-x,t)}{2} \right)}{t^\nu} \, .
        \end{align}
        \item[(c)] Localized initial condition, excited modes, $\I^*(z)$: in the same setting as before, where now we refer only to the time-dependent part, that is, we remove the contribution from the ground state. Directly, we have $P^*(x,t) \equiv P(x,t) - P_\mathrm{eq}(x)$. For $z<z_c$ we have $\mathcal{I}_\mathrm{cont}(z)$, while for $z>z_c$ we have $\mathcal{I}_\mathrm{BG}(z)$. The latter is to compensate for the contribution of the ground state in the region where the particles did not have enough time to diffuse. Explicitly, the rate function for this PDF is written as
        \begin{align}
            \I^*(z) = - \lim\limits_{\stackrel{t,x\to\infty}{z=x/t^\gamma \textrm{fixed}}} \hspace{-0.1in}  \frac{\ln \left( P(x,t) - P_\mathrm{eq}(x) \right)}{t^\nu} \, .
        \end{align}
        \item[(d)] Localized initial condition, odd part, $\I_\mathrm{odd}(z)$: the last possible combination represents the odd part of the PDF, $P_\mathrm{odd}(x,t) = (P(x,t) - P(-x,t))/2$, with localized initial condition. This case is described by $\mathcal{I}_\mathrm{cont}(z)$ for all $z$. Explicitly, the rate function for this PDF is written as
        \begin{align}
            \I_\mathrm{odd}(z) = - \lim\limits_{\stackrel{t,x\to\infty}{z=x/t^\gamma \textrm{fixed}}} \hspace{-0.1in}  \frac{\ln \left( \frac{P(x,t) - P(-x,t)}{2} \right)}{t^\nu} \, .
        \end{align}
    \end{itemize}
    All of these functions are shown in Fig.\,\ref{fig:rate-functions} and in Table\,\ref{tab:scaling-rate-functions} summarizes all possible rate functions and their relationship with the PDF. We cannot extract any more information through the scaling approach. In order to obtain a complete expression for $P(x,t)$, which includes pre-exponential factors, we must make use of the eigenfunction expansion.

    \begin{table}[]
        \centering

        \begin{tabular}{|c|c|c|c|c|}
    \hline
    \begin{tabular}{@{}c@{}}Type \\ of PDF\end{tabular} & $P \to \I$ & \begin{tabular}{@{}c@{}}Rate \\ function \end{tabular} & $\I \to P$ & \begin{tabular}{@{}c@{}}Initial \\ condition\end{tabular} \\
    \hline
    $P_\mathrm{eq}(x)$ & $ \I_\mathrm{eq}(z) = - \lim\limits_{\stackrel{t,x\to\infty}{z=x/t^\gamma \textrm{fixed}}} \hspace{-0.1in}  \frac{\ln P_\mathrm{eq}(x,t)}{t^\nu} $ & $\I_\mathrm{BG}(z)$ & $P_\mathrm{eq} \sim e^{-t^\nu \I_\mathrm{eq}(z)}$ & $\frac{e^{-V(x)/k_B T}}{Z}$ \\
    \hline
    $P^*$ & $\I^*(z) = - \lim\limits_{\stackrel{t,x\to\infty}{z=x/t^\gamma \textrm{fixed}}} \hspace{-0.1in}  \frac{\ln [P(x,t) - P_\mathrm{eq}(x)]}{t^\nu} $ & $ \begin{cases} \I_\mathrm{cont}(z) & z < z_c \\ \I_\mathrm{BG} (z) & z > z_c \end{cases}$ & $P^* \sim e^{-t^\nu \I^* (z)}$  & $\delta(x-x_0)$ \\
    \hline
    $P_\mathrm{even}$ &  $\I_\mathrm{even}(z) = - \lim\limits_{\stackrel{t,x\to\infty}{z=x/t^\gamma \textrm{fixed}}} \hspace{-0.1in}  \frac{\ln [P(x,t)/2 + P(-x,t)/2]}{t^\nu} $ & $ \begin{cases} \I_\mathrm{BG}(z) & z < z_c \\ \I_\mathrm{cont} (z) & z > z_c \end{cases}$ & $P_\mathrm{even} \sim e^{-t^\nu \I_\mathrm{even}(z)}$ & $\delta(x-x_0)$ \\
    \hline
    $P_\mathrm{odd}$ &  $\I_\mathrm{odd}(z) = - \lim\limits_{\stackrel{t,x\to\infty}{z=x/t^\gamma \textrm{fixed}}} \hspace{-0.1in}  \frac{\ln [P(x,t)/2 - P(-x,t)/2]}{t^\nu} $ & $\I_\mathrm{cont}(z)$ & $P_\mathrm{odd} \sim e^{-t^\nu \I_\mathrm{odd}(z)}$ & $\delta(x-x_0)$ \\
    \hline
    \end{tabular}

        \caption{All possible rate functions and their connections to their respective PDFs and their initial conditions. This table summarizes the results presented in Sec.\,\ref{sec:scaling-solution}, items (a) to (d).}
        \label{tab:scaling-rate-functions}
    \end{table}
    
    \section{Scaling and the dynamical phase transition \label{sec:scaled-vars}}

    In this Section, we provide a clear physical meaning to the scaling and to the critical transition point $z_c$. First, we will consider localized initial conditions, that is, the particle starts the motion at position $x_0$ with PDF $P(x,0) = \delta(x-x_0)$, $x_0 \neq 0$. The particle then undergoes Brownian motion under the influence of the external potential, as defined in Eq.\,(\ref{eq:model-potential}). We remind the reader that the force $F(x)$, related to that potential is a restoring force, acting to impede the motion of free diffusion. 

    The PDF at a fixed position $x$, as time goes to infinity, will converge to the Boltzmann-Gibbs equilibrium. However, we cannot expect that, for a fixed time $t$, regardless of how large, the PDF has reached equilibrium for all values of $x$. This is a consequence of the fact that the PDF cannot spread faster than what is allowed by free diffusion, which decays exponentially as a Gaussian, that is, $e^{-x^2/4t}$. Given that the equilibrium distribution decays as a power of $\alpha$, $e^{-V_0 x^\alpha/k_B T}$, for $0 < \alpha < 2$, the more restrictive cutoff is the Gaussian free-particle, and we expect it to dominate the behavior at sufficiently large $x$. 

    An important question is: what is sufficiently large in this case? The answer to this question is given by the critical point obtained in the previous Section. At the point $z=z_c$, of $x = x_c(t) = z_c t^\gamma$, the rate function for cases (b) and (c) in the previous Section undergoes a continuous, but non-analytic, change. The case described in item (b) applies to our current problem of localized initial conditions. This is exactly the point where we shift from the bound region, $x<x_c(t)$, where the PDF has converged to the equilibrium expectation, and the free region, $x>x_c(t)$, where the statistics resemble more closely those of a free-particle. 
    
    This non-analytical transition observed in the rate function for cases (b) and (d) (see previous section) is referred to as a dynamical phase transition \cite{Garrahan2009, Touchette2016, Nyawo2018, Smith2022}. This is due to the rate function playing a similar role to thermodynamic potentials in equilibrium statistical physics \cite{Touchette2016, Smith2022}. In our problem, we have a second-order dynamical phase transition, since both $\I(z)$ and $\I^\prime(z)$ are continuous and only $\I^{\prime \prime}(z)$ is discontinuous \cite{Stella2023}.

    \section{Eigenfunction expansion \label{sec:eigenfunction-approach}}

    As we stated before, the scaling approach, albeit simple and straightforward, is limited in the amount of information we are able to glean from its results. For example, we could obtain the scaling exponents $\nu$ and $\gamma$, we also could find an expression for the critical point $z_c$. However, to calculate ensemble averages of observables, we must obtain pre-exponential corrections which we can calculate through the eigenfunction approach. In this Section, we will obtain a formal solution of the FPE \,(\ref{eq:FPE}) via an eigenfunction expansion of the Fokker-Planck operator
    \begin{align}
        \hat{L}_\mathrm{FP}(x) &= \left\{  \frac{\partial^2}{\partial x^2} - \frac{\partial}{\partial x} \left[ \frac{F(x)}{k_B T} \, \,  \right] \right\} \, , \label{eq:FPE-operator}
    \end{align}
    as
    \begin{align}
        P(x,t) &= \sum_{k=0}^\infty a_k \, {P_k(x)} e^{-\lambda_k t} \, , \label{eq:eigenvalue_expansion}
    \end{align} 
    where $\lambda_k$ are the eigenvalues and $a_k$ are coefficients given by the initial conditions \cite{Risken1996}. We can see from the previous equation that the Fokker-Planck operator is not self-adjoint, and therefore we must take care to distinguish right eigenfunctions, $P_k(x)$ and left eigenfunctions $Q_k(x)$, which are solutions to the adjoint FP operator. A simple way to avoid this inconvenience is to map the FPE into a Schr\"odinger equation via the similarity transformation $\psi_k(x) \equiv e^{-V(x)/2k_B T} P_k(x) = e^{V(x)/2k_B T} Q_k(x)$ \cite{Risken1996}. The problem now reads
    \begin{align}
    - \lambda_k P_k(x) = \hat{L}_\mathrm{FP}(x) P_k(x) \longrightarrow \lambda_k \psi_k(x)  = \hat{H}_S(x) \psi_k(x) \, 
    \end{align}
    where the Schr\"odinger operator is defined as $\hat{H}_S(x) \equiv - e^{V(x)/2k_B T} \hat{L}_\mathrm{FP}(x) e^{-V(x)/2k_B T} = - \partial_x^2 + u_s(x)$, with the effective potential
    \begin{align}
        u_S(x) \equiv \frac{V'(x)^2}{4 (k_B T)^2}  - \frac{V''(x)}{2 k_B T} \, . \label{eq:us}
    \end{align}
    Since the Schr\"odinger operator $\hat{H}_S(x)$ is self-adjoint, the eigenfunctions $\psi_k(x)$ are orthogonal. For simplicity, we will define the scaled potential $v(x) \equiv V(x)/k_B T$ and the inverse temperature $\beta \equiv V_0/k_B T$. We imagine placing the system between reflecting walls at $x = \pm L$, so that the spectrum is guaranteed to be discrete. At the end of the calculation, we will take the limit $L \to \infty$.
    
    In the large $x$ limit, the effective potential becomes
    \begin{align}
        u_S(x) \sim \frac{\beta^2 \alpha^2}{4 x^{2-2\alpha}} + \frac{\beta \alpha(1-\alpha)}{2 x^{2-\alpha}} \, . \label{eq:us-large-x}
    \end{align}
    The crucial point here is the large-$x$ behavior of $u_S(x)$, which is $\sim x^{2(\alpha - 1)}$, since the $-v''(x)$ term goes as $x^{\alpha-2}$, which decays faster. Since both terms decay to zero at large $x$, we essentially have an unbound particle, and we have that the eigenvalues are $\lambda_k \propto 1/L^2$. In the infinite-$L$ limit, the eigenvalue spectrum becomes a continuum that extends down to $\lambda = 0^+$. This is what differentiates this problem from a ``classic" $\alpha > 1$ problem, where $u_S (x)$ grows with $x$ and so the spectrum is discrete near $0$, even in the limit $L \to \infty$.
	
\begin{figure}
        \centering
        \includegraphics[width=0.75\textwidth]{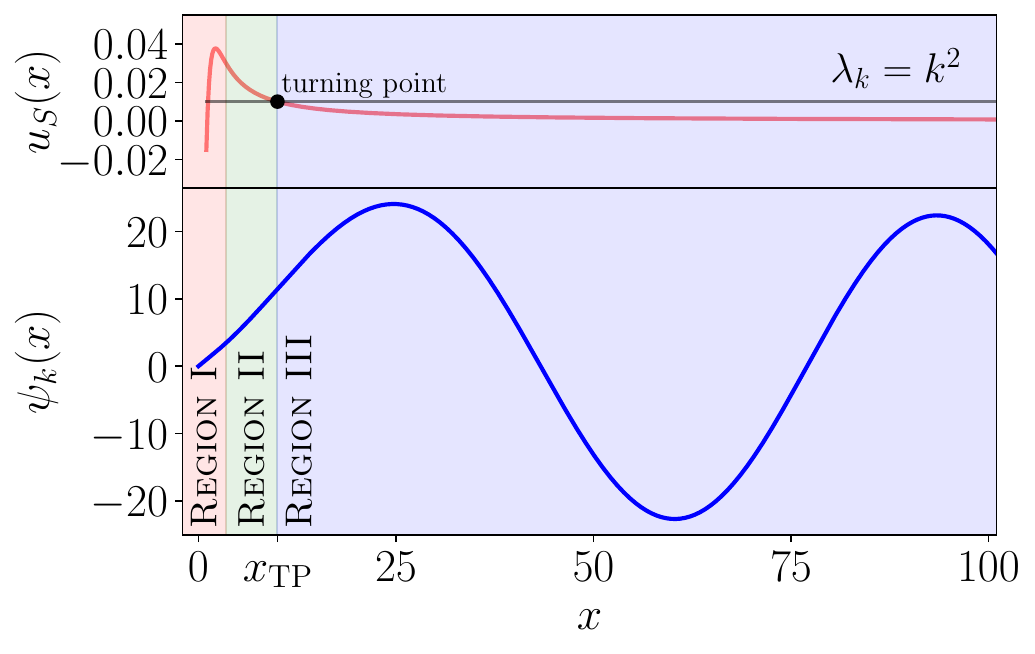}
        \caption{A representation of the different Regions compared to the effective Schr\"odinger potential on the top. In Region I, we have the ``zeroth" approximation, as $u_S(x) \gg \lambda_k$. In Region II, the potential varies smoothly, and the WKB approximation can be used. Note that the turning point $x_{\mathrm{TP}}$ ($u_S(x_\mathrm{TP}) = \lambda_k = k^2$), marks the transition between the exponential growth near the origin and the oscillating, almost free-particle, behavior for large $x$. This last region, which will resemble closely a free-particle, is Region III. We highlight that there are actually two turning points, however, the first turning point is inside Region I, which we describe through the perturbative solutions (see Eqs. (\ref{eq:region-even-1}) and (\ref{eq:region-odd-1})).}
        \label{fig:regions}
    \end{figure}

    Since we are particularly interested in the long-time regime, as we see from Eq.\,(\ref{eq:eigenvalue_expansion}), only the small $\lambda_k$ states contribute to the sum. Thus, the task at hand is to find the continuum state at energy $\lambda_k$, which we will now represent as $\lambda_k \equiv k^2$. We highlight that, since the ground state is $\lambda_k = 0$, all the excited states that form the continuum must be positive and no negative eigenvalue solutions exist. It is also important to highlight that, since we are using scaled dimensionless variables for position and time, the eigenvalue spectrum $\lambda_k$ is also dimensionless. For completeness, in physical units, the relationship between the dimensionless scale and the physical one is $\lambda_k \to \ell^2 k^2$. To accomplish this, we have to solve the problem in 3 different regimes of $x$ and perform asymptotic matching. We will also focus our derivation on $x>0$, as the solutions will either be symmetric (even) or anti-symmetric (odd). The calculations are straightforward, but as we must use several strategies over different scales, we outline the entire procedure
    \begin{itemize}
        \item Since the effective potential $u_S(x)$ is even, the parity is obtained by imposing $\psi^\mathrm{even}(0) = 1$, ${\psi^\mathrm{even}}'(0) = 0 $ for the even eigenfunction, and $ \psi^\mathrm{odd}(0) = 0$ and ${\psi^\mathrm{odd}}'(0) = 1$ for the odd eigenfunction.
        \item We perform a small $k$ expansion in the region near the origin. This solution is valid only for $x$'s which are not too large, in particular, $kx^{\alpha-1}\ll 1$. This range of $x$ values is called Region I. 

        \item In order to extend our solution to larger values of $x$, we will now make use of a WKB solution. We can see in Fig.\,\ref{fig:regions} that there are two turning points, although only one of them (the rightmost one) is highlighted. The first turning point is not important to our analysis, as it is found in Region I for any small $k$ value, and we will only make use of the WKB method outside of that region. The second turning point, which is the relevant one, is the one we highlight (see Fig.\,\ref{fig:regions}).
        
        \item The solution near the origin is then matched with the WKB solution on Region II, to the left of the relevant turning point.
        
        \item We use the standard matching of eigenfunctions in the vicinity of the turning point to get to the right side of the relevant turning point.
        \item For larger $x$, beyond the turning point, we use the oscillating WKB solution, in agreement with the expected free-particle behavior. This is unlike the non-oscillating solution found in Region II.
        \item The eigenvalue spectrum is obtained by imposing the boundary conditions.
        \item Finally, we obtain the value of $a_k$ (see Eq.\,(\ref{eq:eigenvalue_expansion})) through the initial conditions. We point out that $a_k$ ensures the normalization of $P(x,t)$.
    \end{itemize}
    The many regions are outlined in Fig.\,\ref{fig:regions}.

    \subsection{Region I}
\begin{figure}
        \centering
        \includegraphics[width=\textwidth]{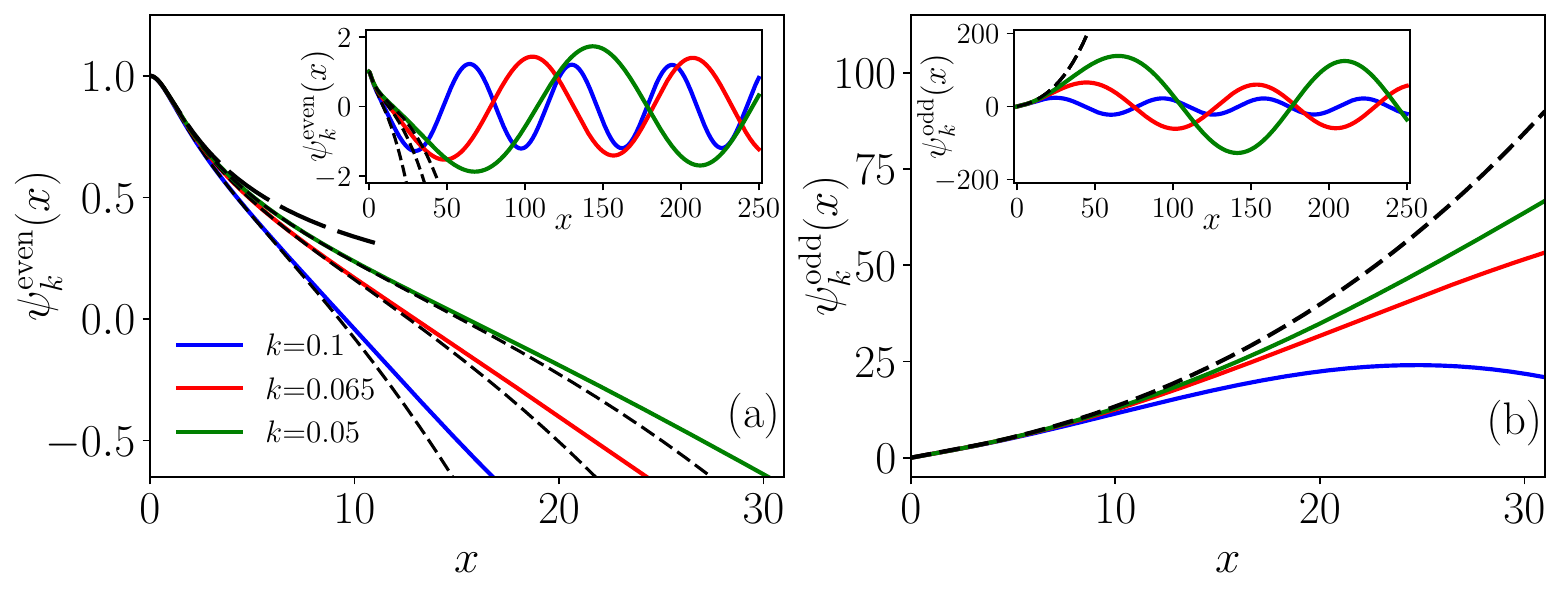}
        \caption{Even (a) and odd (b) eigenfunctions compared to the Region I approximation in Eqs.\,(\ref{eq:region-even-1}) and (\ref{eq:region-odd-1}) (black short-dashed lines) for different eigenvalues (indicated in the legend of panel (a)). On the inset, we show the large $x$ behavior. In panel (a), we also show the Boltzmann-Gibbs ground state in Eq.\,(\ref{eq:BG-ground-state}) (black long-dashed lines). We have taken $\alpha = 1/2$ and $\beta=1$.}
        \label{fig:region-1}
    \end{figure}

     In Region I,  $kx^\alpha \ll 1$,  the $k^2$ term can be treated as a perturbation. The even solutions are obtained by setting ${\psi_k^\mathrm{even}}(0) = 1$. To leading order, all even solutions are approximately the ground state,
    \begin{align}
        \psi_0^\mathrm{even}(x) = e^{-v(x)/2 + v(0)/2} \, . \label{eq:BG-ground-state}
    \end{align}
    This solution is equivalent to the Boltzmann-Gibbs solution in Eq.\,(\ref{eq:Peq}). For $k>0$, we can calculate $\psi_k^\mathrm{even}$ to order $k^2$:
    \begin{equation}
        \psi_k^\mathrm{even}(x)  \approx   e^{-v(x)/2 + v(0)/2} \left[1 - k^2 \int_0^x dy_1 e^{v(y_1)} \int_0^{y_1} dy_2 e^{- v(y_2)} \right] \, . \label{eq:region-even-1}  
    \end{equation}
    At large $x$, the inner integral can be approximated asymptotically as
    \begin{align}
        \int_0^{y_1} e^{- v(y_2)} dy_2 = \int_0^\infty e^{- v(y_2)} dy_2 - \int_{y_1}^\infty e^{- v(y_2)} dy_2 \approx \frac{Z}{2} - \frac{{y_1}^{1-\alpha}}{\beta \alpha} e^{-\beta x^\alpha} \, ,
    \end{align}
    where $Z$ is the Boltzmann-Gibbs partition function, as in Eq.\,(\ref{eq:Peq}).
    The outer integral is dominated by the large $y_1$ region, for which the second term of the inner integral becomes negligible. The asymptotics of the outer integral is then
    \begin{align}
        \frac{Z}{2} \int_0^x e^{v(y)} dy \approx \frac{Z}{2}\left[ \frac{x^{1-\alpha}}{\beta \alpha} - \frac{(1-\alpha)}{\beta^2 \alpha^2} x^{1-2\alpha} \right] e^{\beta x^\alpha} \, .
    \end{align}
    We can now write that the asymptotic expression of the even eigenfunctions is
    \begin{align}
         \psi_k^\mathrm{even}(x) & \approx   e^{v(x)/2 + v(0)/2} \left[1 - \frac{Z k^2}{2} \left( \frac{x^{1-\alpha}}{\beta \alpha} - \frac{(1-\alpha)}{\beta^2 \alpha^2} x^{1-2\alpha} \right) e^{\beta x^\alpha} \right] \nonumber \\
         & \approx   - \frac{Z k^2}{2} e^{\beta x^\alpha /2 + v(0)/2}  \left[ \frac{x^{1-\alpha}}{\beta \alpha} - \frac{(1-\alpha)}{\beta^2 \alpha^2} x^{1-2\alpha} \right] \, . \label{eq:region-even-asym-1}
    \end{align}
    Note that the perturbation has induced the presence of the exponentially growing mode, which dominates at large $x$.  Subsequent terms in the perturbation theory have additional factors of $(kx^{1-\alpha})^2$, so this first-order perturbation theory is reliable as long as
    $k x^{1-\alpha}\ll 1$.
    The approximate odd solution is dominated by the growing, odd, solution of the homogeneous equation and so the leading order is $k$ independent:
    \begin{align}
        \psi^\mathrm{odd}_k(x) \approx e^{-\frac{v(x)}{2} - \frac{v(0)}{2}} \int_0^x e^{v(y)} dy \label{eq:region-odd-1} \, ,
    \end{align}  
    where we have set the normalization by the condition ${\psi_k^\mathrm{odd}}'(0) = 1$. At large $x$, we use the asymptotic form of the integral to write Eq.\,(\ref{eq:region-odd-1}) as
    \begin{align}
	\psi_k^\mathrm{odd}(x) &\approx e^{-\frac{\beta x^\alpha}{2} - \frac{v(0)}{2} } \left[ \frac{x^{1-a}}{\beta a} e^{\beta x^\alpha} - \frac{1 - \alpha}{\beta^2 \alpha^2} x^{1-2\alpha} e^{\beta x^\alpha} \right] \nonumber \\
		& \approx e^{\frac{\beta x^\alpha}{2} - \frac{v(0)}{2}} \left[ \frac{x^{1-\alpha}}{\beta \alpha} - \frac{1 - \alpha}{\beta^2 \alpha^2} x^{1-2\alpha} \right] \, . \label{eq:region-odd-asym-1}
	\end{align}
     We can see the validity of Eqs.\,(\ref{eq:region-even-1}) and (\ref{eq:region-odd-1}) is shown numerically in Fig.\,\ref{fig:region-1}.

    \subsection{Region II}
    
    The next region, which we label Region II, is characterized by $x \gg 1$, whereas we have seen the growing mode is dominant. In this region, we will use the WKB approximation to find the eigenfunctions, which will be valid until we approach too closely to the turning point. From the definition of $u_{S}(x)$ in Eq.\,(\ref{eq:us}), we see that there are in fact two turning points, that is, $u_S (x_{\mathrm{TP}}) = k^2$. The first turning point occurs for $x \sim O(1)$, and is inside Region I, and so is described by that solution and need not concern us further. Here, we are only interested in the second turning point, $x_{\TP} \sim (\beta \alpha/2 k )^{1/(1-\alpha)}$. This turning point marks the boundary between Regions II and III, which we will treat in the next subsection. 
    
     For the values of $x$ we are interested in, the effective Schr\"odinger potential is larger than the eigenvalue. $u_S(x<x_\mathrm{TP}) > k^2$. We also know that the eigenfunction is growing in that region, so we choose the WKB solution
    \begin{align}
         \psi_k(x) \approx  A_k \frac{ \exp \left[- \int_x^{x_\TP} \sqrt{u_S (y) - k^2} \, dy \right] }{[u_S (x) - k^2]^{1/4}} \, . \label{eq:region-2} \, ,
    \end{align}
    where $A_k$ is a constant. To determine $A_k$, we need to evaluate $\psi_k(x)$ far to the left of the turning point, at some $1 \ll x \ll x_\mathrm{TP}$, and match this solution to the asymptotic solutions in Region I. Note that Eqs.\,(\ref{eq:region-even-asym-1}) and (\ref{eq:region-odd-asym-1}) have the same functional dependency on $x$. As a consequence, in Region II both even and odd eigenfunctions will follow Eq.\,(\ref{eq:region-2}), with the only difference being the specific value of $A_k$.

\begin{figure}
        \centering
        \includegraphics[width=\textwidth]{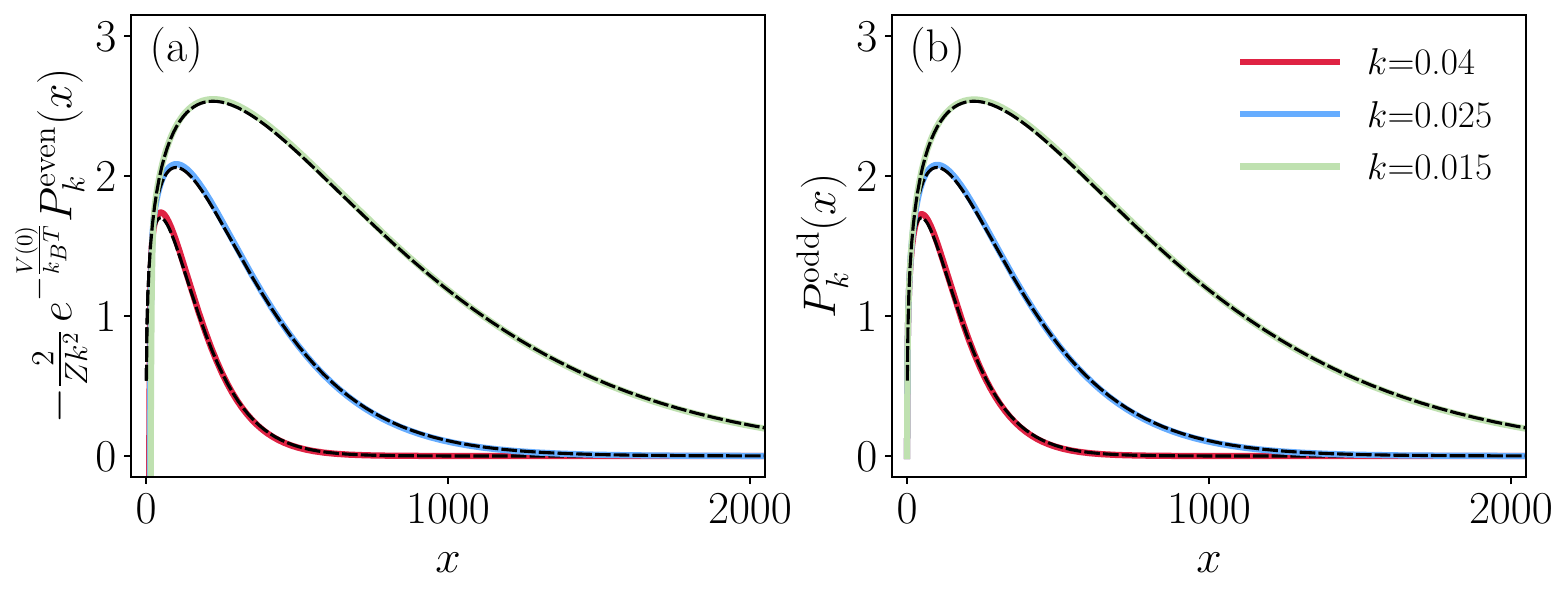}
        \caption{Even (a) and odd (b) right eigenfunctions $P_k(x)$ compared to the Region II approximation in Eqs.\,(\ref{eq:region-even-2}) and (\ref{eq:region-odd-2}) (black short-dashed lines) for different eigenvalues (indicated in the legend of panel (b)). We have taken $\alpha = 3/4$ and $\beta = 1$.}
        \label{fig:Pk-region2}
    \end{figure}

    In the overlap region of interest, $u_S(x) \gg k^2$, we are far to the left of the turning point. This allows us to write the position as a function of the energy $u$,
    \begin{align}
        x(u) & \approx \left( \frac{\beta \alpha}{2 \sqrt{u}} \right)^{1/(1-\alpha)} + \frac{1}{2 \sqrt{u}} \, ,
    \end{align}
    and therefore,
    \begin{align}
        \left| \frac{d x}{d u} \right| & \approx \left( \frac{\beta \alpha}{2} \right)^{1/(1-\alpha)} \frac{u^{\frac{2\alpha - 3}{2 - 2 \alpha}}}{2(1-\alpha)} + \frac{1}{4 u^{3/2}} \, .
    \end{align}
    Let us label the quantity in the exponential of Eq.\,(\ref{eq:region-2}), the WKB action, as $-S(x,k)$. We have that
    \begin{align}
        Q(x) \equiv \int_x^{x_\mathrm{TP}} \sqrt{u_S(y) - k^2}  dy \approx \int_{k^2}^{u_S(x)} \sqrt{u - k^2} \left| \frac{dy}{du} \right| du \, , \label{eq:WKB-Q-action}
    \end{align}
    which we use to split the WKB action into two parts as $Q(x) \approx Q_1(x) + Q_2(x)$, defined from each term of $\left| {d x}/{d u} \right|$ respectively. We expand the first part of the action as
    \begin{align}
		Q_1(x) & \approx \int_{k^2}^{u_S (x)} \sqrt{u - k^2} \left( \frac{\beta \alpha}{2} \right)^{\frac{1}{1-a}} \frac{u^{\frac{2 \alpha - 3}{2-2 \alpha}}}{2(1-\alpha )} du \nonumber \\
		& \approx \frac{ \left( \frac{\beta \alpha}{2} \right)^{\frac{1}{1-\alpha}} }{2(1-\alpha )}\left[ \int_{k^2}^{\infty} \sqrt{u - k^2} \, u^{\frac{2\alpha - 3}{2-2\alpha}} du - \int_{u_S (x)}^{\infty} \sqrt{u - k^2} \, u^{\frac{2\alpha - 3}{2-2\alpha}} du \right] \nonumber \\
		 & \approx \sqrt{\frac{\pi}{4}} \left( \frac{\alpha \beta}{2 k^\alpha} \right)^{\frac{1}{1-\alpha}} \frac{\Gamma\left( \frac{\alpha}{2 - 2 \alpha} \right)}{\Gamma\left( \frac{1}{2 - 2 \alpha} \right)} - \left( \frac{\alpha \beta}{2} \right)^{\frac{1}{1-\alpha}} \left[ \frac{u_S (x)^{-\frac{\alpha}{2-2\alpha}}}{\alpha} \right.  \nonumber \\
        & ~ \left. - \frac{k^2 \, u_S (x)^{-(2-\alpha)/(2-2 \alpha)} }{2 (2- \alpha)} \right] \, .
	\end{align}
    In the limit $x \gg 1$, we can write $u_S(x)$ as in Eq.\,(\ref{eq:us-large-x}), so we may explicitly write
    \begin{align}
        u_S(x)^{-1} & \approx \frac{4 x^{2-2\alpha}}{\beta^2 \alpha^2} \\
        u_S(x)^{-1/2} & \approx \frac{2 x^{1-\alpha}}{\beta \alpha} \left(1 - \frac{1-\alpha}{\beta \alpha} x^{-\alpha} \right) \\
        u_S(x)^{-\alpha / (2-2\alpha)} &\approx  \left( \frac{2}{\beta \alpha} \right)^{\alpha/(1-\alpha)} \left\{ x^\alpha - \frac{1}{\beta} \right\} \, , \\
        u_S(x)^{-(2-\alpha) / (2-2\alpha)} &\approx \left( \frac{2}{\beta \alpha} \right)^{(2-\alpha)/(1-\alpha)} x^{2 - \alpha} \left\{ 1 - \frac{2-\alpha}{\beta \alpha} x^{-\alpha} \right\} \, .
    \end{align}
    Using the previous expressions, we can write that
    \begin{align}
        Q_1(x) \approx  \sqrt{\frac{\pi}{4}} \left( \frac{\alpha \beta}{2 k^\alpha} \right)^{\frac{1}{1-\alpha}} \frac{\Gamma\left( \frac{\alpha}{2 - 2 \alpha} \right)}{\Gamma\left( \frac{1}{2 - 2 \alpha} \right)} - \frac{\beta x^\alpha}{2} + \frac{1}{2} + \frac{k^2 x^{2-\alpha}}{\beta \alpha (2-\alpha)} - \frac{k^2 x^{2-2\alpha}}{\beta^2 \alpha^2} \, ,
    \end{align}
    and
    \begin{align}
      Q_2(x) &= \int_{k^2}^{u_S (x)} \frac{\sqrt{u - k^2}}{4 u^{3/2}}  du = \nonumber \\
     &=  \frac{1}{4} \left[ -2 \sqrt{1 - \frac{k^2}{u_S(x)}} - \ln \left( \frac{k^2}{u_S(x)} \right) + 2 \ln \left( 1 + \sqrt{1 - \frac{k^2}{u_S (x)}} \right) \right] \nonumber \\
      & \approx  - \frac{1}{2} - \frac{1}{4} \ln \left( \frac{k^2}{4 u_S (x)} \right) + \frac{k^2 x^{2-2\alpha}}{2 \beta^2 \alpha^2} \, .
    \end{align}
    Now we will write the same expansion for the WKB denominator, which we will promote to the exponent as
    \begin{align}
        - \frac{1}{4} \ln \left( u_S(x) - k^2 \right) \approx - \frac{1}{4}\ln u_S(x) + \frac{k^2 x^{2-2\alpha}}{\beta^2 \alpha^2}\, .
    \end{align}
    Combining all previous expressions, we reach the complete WKB solution as
    \begin{align}
        \psi_k(x) &\approx A_k \exp \left\{ - Q(x,k) - \frac{1}{4} \ln \left( u_S(x) - k^2 \right)  \right\} \nonumber \\
        &\approx \frac{A_k \, \sqrt{\frac{k}{2}}}{u_S(x)^{-1/2}}  e^{\left\{ - \sqrt{\frac{\pi}{4}} \left( \frac{\alpha \beta}{2 k^\alpha} \right)^{\frac{1}{1-\alpha}} \frac{\Gamma\left( \frac{\alpha}{2 - 2 \alpha} \right)}{\Gamma\left( \frac{1}{2 - 2 \alpha} \right)} + \frac{\beta x^\alpha}{2}  - \frac{k^2 x^{2-\alpha}}{\beta \alpha (2-\alpha)} - \frac{k^2 x^{2-2\alpha}}{\beta^2 \alpha^2} \right\} } \nonumber \\
        \psi_k(x) &\approx A_k \sqrt{2 k} \left[ \frac{x^{1-\alpha}}{\beta \alpha} - \frac{1-\alpha}{\beta^2 \alpha^2} x^{1-2\alpha} \right] e^{\left\{ - \sqrt{\frac{\pi}{4}} \left( \frac{\alpha \beta}{2 k^\alpha} \right)^{\frac{1}{1-\alpha}} \frac{\Gamma\left( \frac{\alpha}{2 - 2 \alpha} \right)}{\Gamma\left( \frac{1}{2 - 2 \alpha} \right)} + \frac{\beta x^\alpha}{2}  - \frac{k^2 x^{2-\alpha}}{\beta \alpha (2-\alpha)} - \frac{k^2 x^{2-2\alpha}}{\beta^2 \alpha^2} \right\} } \nonumber \, . \\ \label{eq:WKB-overlap}
    \end{align}
    As stated in the previous Section, the approximation on Region I is only valid for $x^{2-2\alpha}k^2 \ll 1$. In this limit, the $k^2$ terms in the exponent of Eq.\,(\ref{eq:WKB-overlap}) can be neglected and we see that the WKB solution has the same functional dependence on $x$ that in Eqs.\,(\ref{eq:region-even-asym-1}) and (\ref{eq:region-odd-asym-1}). The constant $A_k$ is finally obtained, for both odd and even eigenfunctions, by matching these equations. We reach that
    \begin{align}
        A_k^{\mathrm{odd}} &= \frac{1}{\sqrt{2k}} \exp \left\{ \sqrt{\frac{\pi}{4}} \left( \frac{\alpha \beta}{2} \right)^{\frac{1}{1-\alpha}} \frac{\Gamma\left( \frac{\alpha}{2 - 2 \alpha} \right)}{\Gamma\left( \frac{1}{2 - 2 \alpha} \right)} \, k^{-\frac{\alpha}{1-\alpha}} - \frac{v(0)}{2} \right\} \label{eq:Ak-odd} \\
        A_k^{\mathrm{even}} &= - \frac{Z \, k^{3/2}}{{2}^{3/2}} \exp \left\{ \sqrt{\frac{\pi}{4}} \left( \frac{\alpha \beta}{2} \right)^{\frac{1}{1-\alpha}} \frac{\Gamma\left( \frac{\alpha}{2 - 2 \alpha} \right)}{\Gamma\left( \frac{1}{2 - 2 \alpha} \right)} \, k^{-\frac{\alpha}{1-\alpha}} + \frac{v(0)}{2} \right\} \label{eq:Ak-even} \, ,
    \end{align}
    and the WKB solutions, far to the left of the turning point, are
    \begin{align}
        \psi_k^{\mathrm{odd}}(x) &=  \left[ \frac{x^{1-\alpha}}{\beta \alpha} -  \frac{1-\alpha}{\beta^2 \alpha^2}  x^{1-2\alpha} \right] e^{\left\{\frac{\beta x^\alpha}{2} - \frac{v(0)}{2} - \frac{k^2 x^{2-\alpha}}{\beta \alpha (2- \alpha)} + \frac{k^2 x^{2-2\alpha}}{2 \beta^2 \alpha^2} \right\} } \label{eq:region-odd-2} \\
       \psi_k^{\mathrm{even}}(x) &= - \frac{Z \, k^2}{2} \left[ \frac{x^{1-\alpha}}{\beta \alpha} -  \frac{1-\alpha}{\beta^2 \alpha^2}  x^{1-2\alpha} \right] e^{\left\{ \frac{\beta x^\alpha}{2} + \frac{v(0)}{2} - \frac{k^2 x^{2-\alpha}}{\beta \alpha (2-\alpha)} + \frac{k^2 x^{2-2\alpha}}{2 \beta^2 \alpha^2} \right\} }  . \label{eq:region-even-2}
    \end{align}
    In Fig.\,\ref{fig:Pk-region2} we show numerically the validity of the previous equations.

    \subsection{Region III}

    The next and final region, which we label Region III, is the region to the right of the turning point. Here, the effective potential is less than the energy eigenvalue, $u_S(x) < k^2$, and the WKB solution is oscillating, as
    \begin{align}
        \psi_k(x) \approx 2 A_k \frac{\sin \left( \int_{x_\TP}^x \sqrt{k^2 - u_S(y) } \, dy  + \frac{\pi}{4} \right)}{(k^2 - u_S(x))^{1/4}} \, , \label{eq:WKB-right}
    \end{align}
    where the constant $2 A_k$ and the phase $\pi/4$ are obtained through the WKB connection formula to cross the turning point. 

    Since our system is placed inside reflective walls at $x=\pm L$, we know that for these points, Eq.\,(\ref{eq:WKB-right}) must follow the no-flux boundary condition. The crucial property in this case is that the difference between the phases of two consecutive eigenvalues $k$ and $k+\Delta k$ must be equal to $\pi$. In the limit of large $L$, specifically $L \gg x_\mathrm{TP}$, we have
    \begin{align}
        \Delta k \int_{x_\TP}^L \frac{k}{\sqrt{k^2 - u_S(y) }} \, dy \approx  L \, \Delta k \approx \pi \, .
    \end{align}
    Clearly, the density of states is continuous, as the spacing $\Delta k \approx \pi/L$ goes to zero as $L \to \infty$, and it stretches from $k=0^+$ to infinity. The complete eigenvalue spectrum is mixed, as we have the discrete ground state for $k=0$, and the continuum states with $k>0$. In the limit of $L \to \infty$, we can replace the sum in $k$ by an integral as
        \begin{align}
        \sum_k \to \int L \frac{dk}{\pi} \, .
    \end{align}

    \subsection{Initial conditions}

    The last step to complete the eigenfunction expansion is to obtain the constants $a_k$ in Eq.\,(\ref{eq:eigenvalue_expansion}). We use the initial condition $P(x,0) = \delta(x-x_0) = \sum_k a_k e^{- v(x)/2} \psi_k(x)$ and the orthogonality of the eigenfunctions $\psi_k(x)$ to write
    \begin{align}
    a_k &= \frac{\int_{-\infty}^{\infty} e^{v(x)/2} P(x,0) \psi_k(x) dx}{\int_{-\infty}^\infty (\psi_k(x))^2 dx} = \frac{ e^{v(x_0)/2} \psi_k(x_0)}{\int_{-\infty}^\infty (\psi_k(x))^2 dx}  \, .
    \end{align}
    The largest contribution to the integral on the denominator arises from Region III. This can be seen in Figs.\,\ref{fig:regions} and \ref{fig:region-1}. In this region, the function behaves as a squared sine function, which we can replace by its mean value 1/2, to find
    \begin{align}
        \int_{-L}^L (\psi_k(x))^2 dx \approx \frac{4 A_k^2}{k} L \, .
    \end{align}
    The previous equation is true for both odd, see Eq.\,(\ref{eq:Ak-odd}), and even, see Eq.\,(\ref{eq:Ak-even}) eigenfunctions. For simplicity, we will take the limit of $x_0 \to 0$, so that we may write $\psi^\mathrm{even}_k(x_0) \approx {\psi^\mathrm{odd}_k}'(x_0)/x_0 \approx 1$.
    
    Finally, the odd PDF can be written as
    \begin{align}
        P^{\mathrm{odd}} (x,t) &= \int_{0}^\infty \frac{dk}{\pi} \frac{k}{4 {A^{\mathrm{odd}}_k}^2} e^{-\frac{v(x)}{2} + \frac{v(x_0)}{2}} \psi_k(x_0) \psi_k(x) e^{-k^2 t} \nonumber \\
       &= x_0 \int_{0}^\infty \frac{dk}{\pi} \frac{k}{4 {A^{\mathrm{odd}}_k}^2} e^{-\frac{v(x)}{2} + \frac{v(0)}{2}} \psi_k(x) e^{-k^2 t} \, ,\label{eq:P_integral_final}
    \end{align}
    and the even PDF can be written as
    \begin{align}
        P^{\mathrm{even}} (x,t) &= \frac{e^{-v(x)}}{Z} + \int_{0}^\infty \frac{dk}{\pi} \frac{k}{4 {A^{\mathrm{even}}_k}^2} e^{-\frac{v(x)}{2} + \frac{v(x_0)}{2}} \psi_k(x_0) \psi_k(x) e^{-k^2 t} \nonumber \\
        &= \frac{e^{-v(x)}}{Z} + \int_{0}^\infty \frac{dk}{\pi} \frac{k}{4 {A^{\mathrm{even}}_k}^2} e^{-\frac{v(x)}{2} + \frac{v(x_0)}{2}} \psi_k(x_0) \psi_k(x) e^{-k^2 t}\, . \label{eq:P_integral_final_even}
    \end{align}

    \section{The equivalence between the eigenfunctions and the scaled rate function approach \label{scaled-rate-eigen}}

    The eigenfunction solution provides a more complete solution for the PDF, compared to the simpler scaling approach used in Sec.\,\ref{sec:scaling-solution}. However, it is not clear from expressions such as Eq.\,(\ref{eq:P_integral_final}), that the results from the scaling approach would emerge. This section is dedicated to showing the equivalence of both approaches.

    First, we must introduce scaled variables for the position $z \equiv x/t^\gamma$ and for the eigenvalues $\kappa \equiv k\, t^\zeta$. The scaling exponents $\gamma$ and $\zeta$ must be obtained. From Eqs.\,(\ref{eq:P_integral_final}) and \,(\ref{eq:P_integral_final_even}), in the regime of small $k$ and restricting the values of $x$ where the WKB solution is valid, we can define the $x$ and $k$ contributions in the exponent as an action, first for the left of the turning point,
    \begin{align}
        S_L(x,k) = k^2 t + \frac{v(x)}{2} + \int_0^{x_\TP} \sqrt{u_S(y) - k^2} \, dy + \int_x^{x_\TP} \sqrt{u_S(y) - k^2} \, dy \, ,
    \end{align}
    and the solution right of the turning point,
    \begin{align}
        S_R(x,k) = k^2 t + \frac{v(x)}{2} + \int_0^{x_\TP} \sqrt{u_S(y) - k^2} \, dy + i \int_x^{x_\TP} \sqrt{k^2 - u_S(y)} \, dy \, .
    \end{align}
    Clearly, the actions defined in the previous equation depend on time, but we omitted $t$ in the argument list for simplicity.
    We take the first two terms of the previous equations, the eigenvalue, and the potential (divided by 2) contribution, and write them using the scaled variables
    \begin{align}
        \frac{\beta x^\alpha}{2} &= \frac{\beta z^\alpha}{2} t^{\alpha \gamma} \label{eq:pot-half-scaling} \\
        k^2 t &= \kappa^2 t^{1 - 2\zeta} \label{eq:k2t-scaling} \, .
    \end{align}
    At this point, we will impose that the scaling exponents must ensure that all relevant terms, i.e. terms that depend on $x$ and $k$, contribute equally to the action. We thus obtain that $1 - 2\zeta = \alpha \gamma$. Next, we must ensure that the Schr\"odinger potential and the eigenvalues have the same dependency in time, that is
    \begin{align}
        u_S(x) &= \frac{\alpha^2 \beta^2}{4 x^{2-2\alpha}} + \frac{\beta \alpha(1-\alpha)}{2 x^{2-\alpha}} \to \frac{\alpha^2 \beta^2}{4 z^{2-2\alpha}} t^{\gamma(\alpha-1)} \label{eq:us-scaling} \\
        k^2 &= \kappa^2 t^{-2 \zeta} \label{eq:k2-scaling} \, ,
    \end{align}
    where we neglected the second term in $u_S(x)$ as it will decay in time faster than the first. Matching the time-dependence in Eqs.\,(\ref{eq:pot-half-scaling}), \,(\ref{eq:k2t-scaling}), \,(\ref{eq:us-scaling}), and\,(\ref{eq:k2-scaling}), we find that the scaling exponents must satisfy
    \begin{align}
         1 - 2 \zeta = \alpha \gamma ~~ \mathrm{and} ~~ 2 \zeta = \gamma (1 - \alpha) \, ,
    \end{align}
    which we can solve to recover $\gamma$ in Eq.\,(\ref{eq:scaled-exponents}) and find
    \begin{align}
        \zeta  = \frac{1 - \alpha}{2 - \alpha} \, .
    \end{align}
    The second exponent $\nu$ is found through
    \begin{align}
        \frac{v(x)}{2} + k^2 t &\to t^{-\frac{\alpha}{2-\alpha}} \left[ \frac{\beta z^\alpha}{2} + \kappa^2 \right] = t^{-\nu} \left[ \frac{\beta z^\alpha}{2} + \kappa^2 \right] \,.
    \end{align}
    Region I is defined by $k^2 x^{2-2\alpha} \ll 1$. In the scaled variables, we have that $\kappa^2 z^{2-2\alpha} t^\nu \ll 1$. Therefore, in the limit of large $t$, Region I shrinks and Regions II and III, which follow the WKB approximation, extend for all values of $z$. We also highlight that the WKB approximation becomes more accurate as time increases, becoming exact in the limit $t \to \infty$.
    
\begin{figure}
        \centering
        \includegraphics[width=0.75\textwidth]{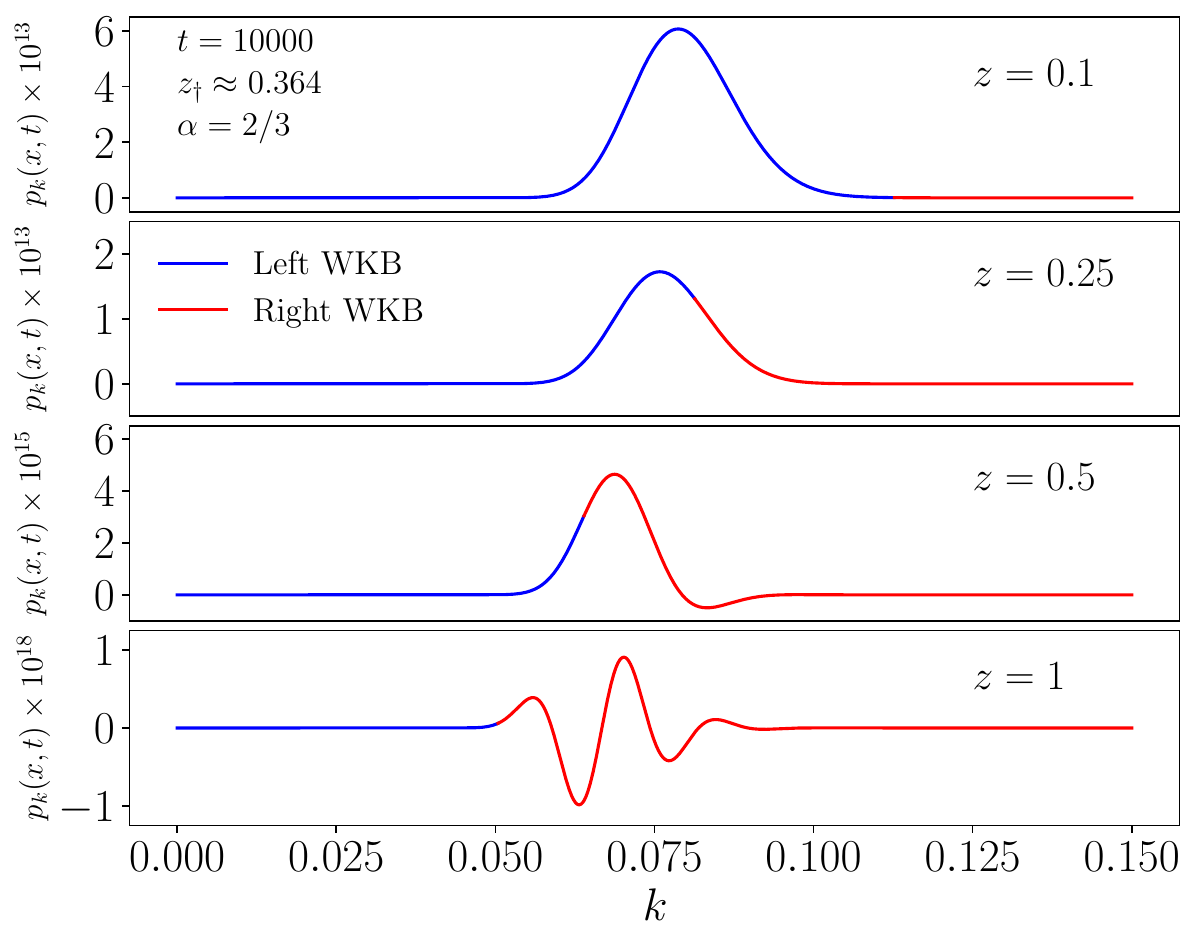}
        \caption{The $k$-density $p_k(x,t)$, Eq.\,(\ref{eq:k-density}), for different values of the scaled variable $z$, versus the eigenvalue $k$. The contribution from the left WKB solution is shown in blue and the right is shown in red. For $z<z_\dagger$, we can see clearly that the main contribution comes from the left WKB. For this range, we have a clear maximal point. On the other hand, for $z>z_\dagger$, the main contribution comes instead from the right WKB solution, which is highly oscillating. The time is fixed $t = 10000$, $\beta=1$, and $\alpha = 2/3$.}
        \label{fig:pk-density}
    \end{figure}

    Now that we already recovered the exponents, we need to show that the rate function, as defined in Eq.\,(\ref{eq:scaled-exponents}), can be extracted from the eigenfunction approach. Let us re-write Eq.\,(\ref{eq:P_integral_final}) as
    \begin{align}
        P_\mathrm{odd} (x,t) = \int_{0}^{\infty} dk \, p_k(x,t) \, , \label{eq:k-density}
    \end{align}
    where we introduce the $k$-density $p_k(x,t)$. This can also be done for the excited, time-dependent, part of Eq.\,(\ref{eq:P_integral_final_even}). In order to find the PDF at position $x$ and time $t$, we must integrate the $k$-density $p_k(x,t)$ over all values of $k$. For each value of $x$, we will also have a turning point $k_\TP = \sqrt{u_S(x)}$, so that for $k<k_\TP$ we must use the left WKB solution and $k<k_\TP$ we must use the right WKB solution. The same is true in the scaled variables, with the turning point energy $\kappa_\TP = \beta \alpha / (2 z^{1-\alpha})$. We have
    \begin{align}
        S(x,k) \to t^\nu \mathcal{S}(z,\kappa) \, ,
    \end{align}
    with the left solution, $\kappa < \kappa_\TP$,
    \begin{align}
        \mathcal{S}_L(z,\kappa) &= \kappa^2 + \frac{\beta z^\alpha}{2} + \int_0^{z_\TP} \sqrt{\frac{\beta^2 \alpha^2}{4 y^{2-2\alpha}} - \kappa^2} \, dy + \int_z^{z_\TP} \sqrt{\frac{\beta^2 \alpha^2}{4 y^{2-2\alpha}} - \kappa^2} \, dy \, , \label{eq:SL-scaled}
    \end{align}
    and right solution, $\kappa > \kappa_\TP$,
    \begin{align}
        \mathcal{S}_R(z,\kappa) &= \kappa^2 + \frac{\beta z^\alpha}{2} + \int_0^{z_\TP} \sqrt{\frac{\beta^2 \alpha^2}{4 y^{2-2\alpha}} - \kappa^2} \, dy + i \int_z^{z_\TP} \sqrt{\kappa^2 - \frac{\beta^2 \alpha^2}{4 y^{2-2\alpha}}} \, dy \, .
    \end{align}
    In the two previous equations, we have used that
    \begin{align}
        \int_0^{z_\TP} \sqrt{\frac{\beta^2 \alpha^2}{4 y^{2-2\alpha}} - \kappa^2} \, dy = \sqrt{\frac{\pi}{4}} \left( \frac{\alpha \beta}{2 } \right)^{\frac{1}{1-\alpha}} \frac{\Gamma\left( \frac{\alpha}{2 - 2 \alpha} \right)}{\Gamma\left( \frac{1}{2 - 2 \alpha} \right)} \kappa^{-\frac{\alpha}{1-\alpha}} \, ,
    \end{align}
    to simplify the expression. Remarkably, this term is also present in the context of the rate function for the time average of the $\alpha^\textit{th}$ power of the trajectory of a standard Ornstein–Uhlenbeck process \cite{Touchette2018,Smith2022b}.
    From Fig.\,\ref{fig:pk-density} it becomes clear, at least for small $z$, that the largest contribution to the integration comes from the vicinity of a specific eigenvalue $k^*$, or $\kappa^*$. The exponent of the $k$-density is $S(x,k)$, and the critical points are found through $\partial_k S(x,k)|_{k=k^*} = 0$, and, in the scaled coordinates
    \begin{align}
        \left. \frac{\partial}{\partial \kappa} \mathcal{S}(z,\kappa) \right|_{\kappa=\kappa^*} = 0 \, . \label{eq:critical-kappa}
    \end{align}

    For now, we will focus on solving the odd part of the PDF. This will allow us to calculate the integral without concerning ourselves with the Boltzmann-Gibbs ground state. The integral of the $k$-density can be approximated using Laplace's method and we can write, up to pre-exponential terms, that
    \begin{align}
        P_\mathrm{odd}(x,t) \approx e^{-t^\nu \mathcal{S}(z,\kappa^*)} \, .
    \end{align}
    The eigenfunction expansion and the rate function scaling approach must yield the same expression for the PDF, that is,
    \begin{align}
        \I(z) = \mathcal{S}(z,\kappa^*) \, . \label{eq:equiv-action-wkb}
    \end{align}
    A direct way to show that this is true is to demonstrate that $\mathcal{S}(z,\kappa^*)$ is also a solution of the differential Eq.\,(\ref{eq:df-rate-function}). Note that
    \begin{align}
        \frac{\partial}{\partial z} \mathcal{S}(z,\kappa^*) = \left. \frac{\partial}{\partial z} \mathcal{S}(z,\kappa) \right|_{\kappa = \kappa^*} + \left. \frac{\partial}{\partial \kappa} \mathcal{S}(z,\kappa) 
        \right|_{\kappa  = \kappa^*} \frac{\partial \kappa^*}{\partial z} = \left. \frac{\partial}{\partial z} \mathcal{S}(z,\kappa) \right|_{\kappa = \kappa^*} \, ,
    \end{align}
    where the last equality is true due to $\kappa^*$ being the point of maximal contribution. We will now show that Eq.\,(\ref{eq:equiv-action-wkb}) is true. One important straightforward relation is that,
    \begin{align}
    \frac{\partial}{\partial z} \left[ \frac{z}{\alpha} \sqrt{\frac{\beta^2 \alpha^2}{4 z^{2-2\alpha}} - \kappa^2} \right] = \sqrt{\frac{\beta^2 \alpha^2}{4 z^{2-2\alpha}} - \kappa^2} - \frac{1-\alpha}{\alpha} \frac{\kappa^2}{\sqrt{\frac{\beta^2 \alpha^2}{4 z^{2-2\alpha}} - \kappa^2}} \, ,
    \end{align}   
    which allows us to write 
    \begin{align}
       \left[ \frac{1-\alpha}{2-\alpha} \right] {\kappa} \frac{ \partial S }{\partial \kappa} &= \kappa^2 + \frac{z \frac{\partial \mathcal{S}}{\partial z} - \alpha \mathcal{S}}{2-\alpha} \, ,
    \end{align}
    and
    \begin{align}
        \left[ \frac{\partial \mathcal{S}}{\partial z}  \right]^2 
        &= \frac{\beta \alpha}{z^{1-\alpha}} \frac{\partial \mathcal{S}}{\partial z} - \kappa^2 \, .
    \end{align}
    Combining both previous equations, we can write the equivalent of Eq.\,(\ref{eq:df-rate-function}) as
    \begin{align}
        \left[ \frac{\partial \mathcal{S}}{\partial z}  \right]^2 - \left( \frac{\beta \alpha}{z^{1-\alpha}} + \frac{z}{2-\alpha} \right) \frac{\partial \mathcal{S}}{\partial z} + \frac{\alpha \mathcal{S}}{2-\alpha} = \left[ \frac{1-\alpha}{2-\alpha} \right] {\kappa} \frac{ \partial S }{\partial \kappa} \, . \label{eq:diff-eq-wkb-action}
    \end{align}
    It is clear that
    $\kappa = \kappa^*$ yields the excited solutions and $\kappa = 0$  is equivalent to the Boltzmann-Gibbs solution. Therefore, we have that
    \begin{align}
        \I_\mathrm{BG}(z) &= \mathcal{S}(z,0) \, ,
    \end{align}
    and
    \begin{align}
        \I_\mathrm{cont}(z) &= \mathcal{S}(z,\kappa^*) \, .
    \end{align}
    In these cases, Eq.\,(\ref{eq:diff-eq-wkb-action}) is identical to Eq.\,(\ref{eq:df-rate-function}), which is sufficient to show the validity of the general Eq.\,(\ref{eq:equiv-action-wkb}).

    Now, our task is to obtain $\kappa^*$ for different values of $z$. It is important to note that, depending on the value of $z$, the maximum contribution may arise from either the left or the right solution. This effect is shown in Fig.\,\ref{fig:pk-density}, where we see that the left solution provides the dominant contribution for small $z$ and the right oscillating contribution dominates for large $z$.

    \subsection{Examples}

    An important example is the $z=0$ case, where we know the maximum contribution comes from the left solution. In that case, we find ${\kappa^*}$ through,
    \begin{align}
        {\kappa^*}^2 + \frac{\alpha}{2-\alpha} \mathcal{S}_L(0,{\kappa^*}) &= 0 \, ,
    \end{align}
    as
    \begin{align}
        {\kappa^*}(0) = \left[ \sqrt{\frac{\pi}{4}} \left( \frac{\beta \alpha}{2} \right)^{\frac{1}{1-\alpha}} \frac{\Gamma\left( \frac{\alpha}{2 - 2 \alpha} \right)}{\Gamma\left( \frac{1}{2 - 2 \alpha} \right)} \, \frac{\alpha}{1-\alpha} \right]^{\frac{1-\alpha}{2-\alpha}} \, . \label{eq:critical-kappa-z0}
    \end{align}
    The value of the WKB action, which is equivalent to the excited branch of the rate function, at $z=0$, is
    \begin{align}
         \mathcal{S}_L(0,{\kappa^*}) = \I_\mathrm{cont}(0) = \frac{\left( \sqrt{\pi} \left( \frac{\alpha V_0}{2 k_B T} \right)^{\frac{1}{1-\alpha}} \frac{\Gamma\left( \frac{\alpha}{2 - 2 \alpha} \right)}{\Gamma\left( \frac{1}{2 - 2 \alpha} \right)} \right)^{1-\nu}}{\nu^\nu(1-\nu)^{1-\nu}} \, .
    \end{align}

    There is a maximum value of the scaled position, $z_\dagger$, where the main contribution switches from the left WKB function to the right one. This point is the point where the zero derivative point $\kappa^*$ is identical to the turning point $\kappa_\TP$. Also at this point, we have that $z_\dagger = z_\TP$, and therefore
    \begin{align}
        \kappa^*(z_\dagger) =  \left[ \sqrt{\frac{\pi}{16}} \left( \frac{\beta \alpha}{2} \right)^{\frac{1}{1-\alpha}} \frac{\Gamma\left( \frac{\alpha}{2 - 2 \alpha} \right)}{\Gamma\left( \frac{1}{2 - 2 \alpha} \right)} \, \frac{\alpha}{1-\alpha} \right]^{\frac{1-\alpha}{2-\alpha}} \, .
    \end{align}

    For $z>z_\dagger$, the dominant contribution to the integral comes from the right solution. Since this solution is an oscillating function, in order to use Laplace's method, we must perform an analytical continuation of the right function. This allows us to search for the maximum point beyond the range of validity of $\kappa$, so that the problem becomes one of finding the steepest descent trajectory. In that region, we can manipulate the expression for the right solution to find
    \begin{align}
        \mathcal{S}_R(z,\kappa) = \kappa^2 + \frac{\beta z^\alpha}{2} + \int_0^z \sqrt{\frac{\beta^2 \alpha^2}{4 y^{2-2\alpha}} - \kappa^2} dy \, , \label{eq:SR-scaled}
    \end{align}
    and is directly related to the expression for the left solution as
    \begin{align}
        \mathcal{S}_R(z,\kappa) = \mathcal{S}_L(z,\kappa) - 2 \int_0^{z_\TP} \sqrt{\frac{\beta^2 \alpha^2}{4 y^{2-2\alpha}} - \kappa^2} dy \, ,
    \end{align}
    which, as we will see in  Sec. \ref{sec:path-integral} describing the path-integral approach, has a deeper physical meaning.

    We find that, as expected, the maximum for $z = z_\dagger$ is identical for both left and right WKB solutions. For $z> z_\dagger$, the maximum is found for a real value of $\kappa$, albeit in the region where the right solution would not be valid. 

    The integral of the $k$-density is defined in the range $(0,\infty)$. When the right solution provides the dominant contribution, we can re-write this integral approximately as
    \begin{align}
        \int_0^\infty e^{-t^\nu \mathcal{S}(z,\kappa)} d\kappa \approx \int_{\kappa_\TP}^\infty e^{-t^\nu \mathcal{S}_R (z,\kappa)} d\kappa \, . \label{eq:SR-z-larger-dagger}
    \end{align}
    Using the analytical continuation, we define a contour $C$ that goes through the complex plane, ending at $\kappa = \kappa_\TP$ and starting at $\kappa \to \infty$. As there are no poles inside the contour, we can write that
    \begin{align}
        \int_{\kappa_\TP}^\infty e^{-t^\nu \mathcal{S}_R (z,\kappa)} d\kappa + \int_{C} e^{-t^\nu \mathcal{S}_R (z,\kappa)} d\kappa = 0 \, .
    \end{align}
    The last component of the method is to ensure that the contour crosses the critical point $\kappa^*$ to form a local maximum to the integrand, which is equivalent to a minimum of $\mathcal{S}_R(z,\kappa)$. This means that we cross the $\kappa^*$ point, which is a real number, not through the real axis, but at an angle of $\pi/2$ in the complex plane. 
    
    As we approach the $\kappa^*=0$ region, the expression for the right solution can be approximated as,
    \begin{align}
        \mathcal{S}_R(z,0) & \approx \frac{\beta z^\alpha}{2} + \kappa^2 + \int_0^z \left[ \frac{\beta \alpha}{2 y^{1-\alpha}} - \kappa^2 \frac{y^{1-\alpha}}{\beta \alpha} \right] dy \nonumber \\
        & \approx  \beta z^\alpha + \kappa^2 \left( 1 - \frac{z^{2-\alpha}}{\beta \alpha (2-\alpha)} \right) =   \, .
    \end{align}
    The limit case where $z = z_c$, the critical eigenvalue is $\kappa^* = 0$, which is equivalent to the Boltzmann-Gibbs solution,
    \begin{align}
        \mathcal{S}_R(z,0) &= \frac{\beta z^\alpha}{2} + \int_0^{\infty} \sqrt{\frac{\beta^2 \alpha^2}{4 y^{2-2\alpha}}} \, dy + i \int_z^{\infty} \sqrt{- \frac{\beta^2 \alpha^2}{4 y^{2-2\alpha}}} \, dy \, = \beta z^\alpha \,
    \end{align}
    The value of $z_c$ can be recovered using a small $\kappa$ expansion for the right WKB solution, that is,

    For values of the scaled position larger than the critical value, $z>z_c$, the maximum eigenvalues are all pure imaginary numbers. Here, we will change our strategy to calculate the integral. We will make use of the fact that the is symmetric in $\kappa$ to replace the limit of integration to $(-\infty,\infty)$, where we will then close the integration contour crossing the critical point $\kappa^*$. It is critical that we cross this point with a horizontal line (constant imaginary part). When we consider the limit of very large $z$, the critical eigenvalue is obtained as
    \begin{align}
        \mathcal{S}_R(z,\kappa) \approx \kappa^2 + i z \kappa \, ,
    \end{align}
    $\kappa^* = iz/2$ and $\mathcal{S}_R(z,\kappa^*) = z^2/4$, which is identical to the free particle solution.
    
\begin{figure}
        \centering
        \includegraphics[width=0.85\textwidth]{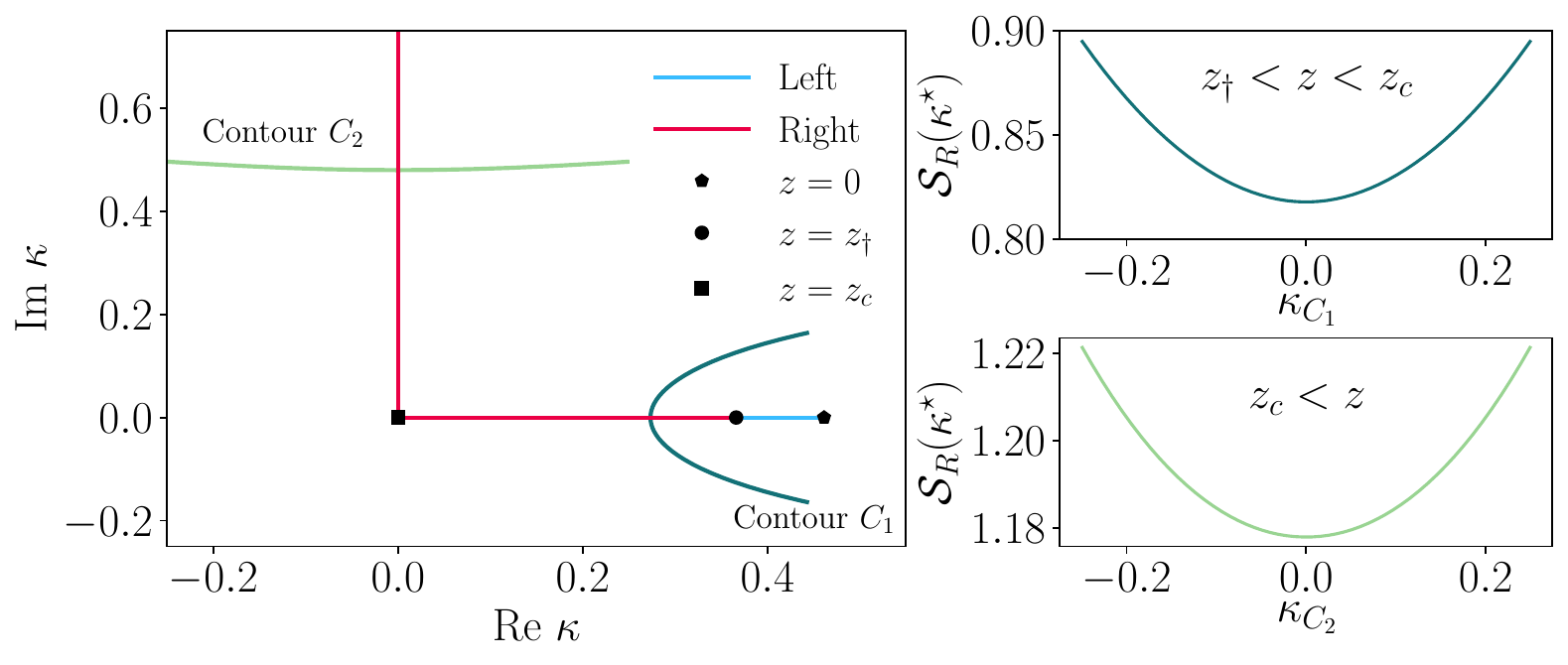}
        \caption{The integration contour of $\kappa$ in the complex plane where all possible critical points are highlighted in the solid colored lines. For the range $0<z<z_\dagger$, the critical point is plotted in light blue, and the maximal contribution comes from the Left WKB function (shown in legend).  In the range $z_\dagger<z<z_c$, the maximal contribution comes from the right WKB solution, and the critical $\kappa^\star$ is found in the real line. Finally, in the last possible range $z>z_c$, all critical $\kappa^\star$ are found in the pure imaginary line. On the right, we show two examples of the contours of the steepest descent. We have used $\alpha = 1/2$ and $\beta = 1$.}
        \label{fig:integration-contour}
    \end{figure}

    In Fig.\,\ref{fig:integration-contour}, we show some examples of contours for values of $z>z_\dagger$. We highlight that we have also numerically compared solutions for the rate function $\I(z)$ obtained through the scaling approach with $\mathcal{S}(z,\kappa^*)$ with complete agreement.

    \section{The solvable $V$-shaped potential ($\alpha=1$) \label{sec:V-shaped}}

    In this section, we shall treat the analytically solvable $V$-shaped potential, which is equivalent to $\alpha=1$ \cite{Chen2013,Majumdar2020,Zarfaty2022}, defined below. As we will see, the eigenvalue spectrum in this case is different than the $0 < \alpha < 1$ case. Here, the spectrum has a gap between the ground state and the beginning of the excited state. Despite that, we can use it to gather valuable insights into the behavior of the solution, as this potential has a simple analytical solution. The $V$-shaped potential is defined as
    \begin{align}
        V(x) &= V_0 |x/\ell| \, .
    \end{align}
    
    \subsection{Using the rate-function approach}

    This potential is equivalent to the $\alpha = 1$ case for our family of potentials, see Eq.\,(\ref{eq:model-potential}). Here, the scaling exponents are $\gamma = \nu = 1$, and the scaled position variable for this example is $z \equiv x / t$. Following the scaling approach, we find that the differential equation for the rate function $\I(z)$ is
    \begin{align}
        \I'(z)^2 - \left( \frac{V_0}{k_B T} + z \right) \I'(z) + \I(z) = 0 \, , \label{eq:def-eq-rate-V}
    \end{align}
    which is equivalent to Eq.\,(\ref{eq:df-rate-function}), for $\alpha = 1$. As we saw, one immediate solution is the Boltzmann-Gibbs equivalent, that is, $\I(z) = V_0 z / k_B T$. From the form of Eq.\,(\ref{eq:def-eq-rate-V}), we can propose a polynomial solution as $\I(z) = b_2 z^2 + b_1 z + b_0$. Plugging that expression into Eq.\,(\ref{eq:def-eq-rate-V}), we find that
    \begin{align}
        b_0 + b_1^2 &= \frac{b_1 V_0}{k_B T} \, \\
        b_1 &= \frac{V_0}{2 k_B T} \, .
    \end{align}
    Imposing diffusive behavior for large $z$, we find that $b_2 = 1/4$, $b_1=0$, and $b_0 = V_0^2 / 4 (k_B T)^2$. The complete solution is
    \begin{align}
        \I_{\mathrm{BG}} (z) &= \frac{V_0}{k_B T} z \\
        \I_{\mathrm{cont}} (z) &= \frac{z^2}{4} + \frac{V_0}{2 k_B T} z + \frac{V_0^2}{4 (k_B T)^2} \, . \label{eq:def-I-cont-V}
    \end{align}
In this particular case, we emphasize that the rate function related to the continuum contributions, $\I_{\mathrm{cont}}$, is also a zero discriminant line. The critical point is found through $\I_{\mathrm{BG}} (z_c) = \I_{\mathrm{cont}}(z_c)$, that is,
\begin{align}
    z_c = \frac{V_0}{k_B T} \, . \label{eq:critical-point-via-rate-function}
\end{align}
Note that, in this case, the scaling exponents are identical to unity, so here this is an exponential (not stretched) relaxation.

\subsection{Using the eigenfunctions approach}

For the $V$-shaped potential, we have that the Schr\"odinger equivalent potential is
\begin{align}
    u_S(x) = \frac{V_0^2}{4 (k_B T)^2} - \frac{V_0}{2 k_B T} \delta(x) \, ,
\end{align}
where $\delta(x)$ is the Dirac delta function. For $x\neq 0$, the Schr\"odinger equation is equivalent to that of a constant potential  (with intensity ${V_0^2}/{4 (k_B T)^2}$). The eigenvalue spectrum has the form
\begin{align}
    \lambda_k = \frac{V_0^2}{4 (k_B T)^2} + k^2 \, . 
\end{align}
 We see here that there is a gap $\frac{V_0^2}{4 (k_B T)^2} $ between the ground state and the first excited state. The complete normalized eigenfunctions are obtained using the Dirac delta function to match the $x>0$ and $x<0$ regions. Explicitly, the normalized even and odd eigenfunctions are
\begin{align}
    \psi_k^\mathrm{even}(x) &= \frac{k \cos(kx) + \left( \frac{V_0}{2 k_B T} \right) \sin(k |x|)}{\sqrt{\pi \left(k^2 + \frac{V_0^2}{4(k_BT)^2} \right)}} \, , \\ 
    \psi_k^{\mathrm{odd}}(x) &= \frac{1}{\sqrt{\pi}} \sin(kx) \, . \label{eq:eigenfunction-V-odd}
\end{align}
To obtain the PDF, we must perform an integral over positive $k$. Let us consider a localized initial condition $P(x,t=0) = \delta(x-x_0)$. For a small initial position $x_0$, we can write $\psi_k(x_0) = \sin(kx_0) / \sqrt{\pi} \approx k x_0 / \sqrt{\pi}$. The odd part of the probability density is written as
\begin{align}
    P_{\mathrm{odd}} (x,t) &=  \int_0^\infty dk \psi^\mathrm{odd}_k(x) \psi^\mathrm{odd}_k(x_0) e^{-\lambda_k t - \frac{V_0 |x|}{2 k_B T}} \nonumber \\ &= e^{- \frac{V_0 |x|}{2 k_B T}} \int_{-\infty}^\infty \frac{dk}{2 \pi}  k x_0 \sin(kx) \exp \left( -k^2 t - \frac{V_0^2 t}{4 (k_B T)^2} \right) \, \nonumber \\
    &= x_0 e^{- \frac{V_0 |x|}{2 k_B T} - \frac{V_0^2 t}{4 (k_B T)^2}} \int_{-\infty}^\infty \frac{dk}{2 \pi i} \exp \left( -k^2 t + i k x  \right) .
\end{align}
The last integral is a very simple Gaussian integral. Despite this, we will go into detail as to how this integral is calculated as it will be useful to understanding the dynamical phase transition in the even PDF. First we introduce the same scaled position variable $z \equiv x/t$, so that
\begin{align}
    P_{\mathrm{odd}} &= x_0  e^{- \frac{V_0 |x|}{2 k_B T} - \frac{V_0^2 t}{4 (k_B T)^2}} \int_{-\infty}^\infty \frac{dk}{2 \pi i} k \exp \left( t [ -k^2  + i k z]  \right) . \label{eq:no-prefactor}
\end{align}
Extending the variable $k$ into the complex plane, we find that the value $k^*$ that maximizes the argument inside the exponential is
\begin{align}
    k^* = \pm \frac{i z}{2} \, .
\end{align}
Since there are no poles in the pre-factor of the integral in Eq.\,(\ref{eq:no-prefactor}), we make use of Cauchy's integral theorem to close the contour with a new path that crosses the maximum point $k^*$. We then use Laplace's method to obtain the integral. Despite that, in the particular case of the $V$-shaped potential, the integral is Gaussian and this result is analytical, we detail the steps to calculate it as we will use them in the next section where we treat $0<\alpha<1$. We write the integration result as
\begin{align}
    P_\mathrm{odd} (x,t) &\approx \frac{x_0 x}{4 \sqrt{\pi} t^{3/2}} \exp \left(- t \left[ \frac{z^2}{4} + \frac{V_0}{2 k_B T} |z| + \frac{V_0^2}{4 (k_B T)^2} \right] \right) \, ,
\end{align}
where we can clearly see that the expression in the exponent is identical to $\I_\mathrm{cont}(z)$ in Eq.\,(\ref{eq:def-I-cont-V}). 

We now turn to the contribution from the continuum modes to the even part of the PDF, $P^*_\mathrm{even}(x,t) = P_\mathrm{even}(x,t) - P_\mathrm{eq}(x)$. In that case, we have
\begin{align}
    P^*_\mathrm{even}(x,t) &= \int_0^\infty dk \psi^\mathrm{even}_k(x) \psi^\mathrm{even}_k(0) e^{-\lambda_k t - \frac{V_0 |x|}{2 k_B T}} \nonumber \\
    &= \int_{-\infty}^\infty \frac{dk}{2 \pi} \frac{k^2 \cos(kx) + k \left( \frac{V_0}{2 k_B T} \right) \sin(k |x|)}{\left(k^2 + \frac{V_0^2}{4(k_BT)^2} \right)} e^{ -k^2 t - \frac{V_0^2}{4 (k_B T)^2}t  - \frac{V_0 |x|}{2 k_B T}} \, , 
\end{align}
where, for simplicity, we take $x_0 = 0$. Note that we could employ Laplace's method to solve the Gaussian integrals in the same way as we did for the odd part. However, unlike the odd part, here the pre-factor contains a pole at $\bar{k} = \pm i V_0/ k_B T$. For a critical value $z_c$ of the scaled variable, we will have that $k^* = \bar{k}$, explicitly
\begin{align}
    \frac{i z_c}{2} = \pm i \frac{V_0}{2 k_B T} \to z_c = \pm \frac{V_0}{ k_B T} \, .
\end{align}
This is the same critical point as obtained in Eq.\,(\ref{eq:critical-point-via-rate-function}) through the rate function approach. For $|z|<|z_c|$, the pole will lie outside the integration contour, and will not contribute to the final result. On the other hand, if $|z|>|z_c|$, we must include the contribution from the residue of $\bar{k}$, which is equivalent to the (negative) Boltzmann-Gibbs solution,
\begin{align}
    (2 \pi i) \mathrm{Res}_{k=\bar{k}} \left\{  \frac{1}{2 \pi} \frac{k^2 \cos(kx) + k \left( \frac{V_0}{2 k_B T} \right) \sin(k |x|)}{\left(k^2 + \frac{V_0^2}{4(k_BT)^2} \right)} e^{ -k^2 t - \frac{V_0^2}{4 (k_B T)^2}t  - \frac{V_0 |x|}{2 k_B T}} \right\} = - \frac{V_0}{2 k_B T} e^{- \frac{V_0 |x|}{k_B T} } \, . \nonumber \\ 
\end{align}
For $z > z_c$, this negative Boltzmann-Gibbs contribution will dominate. In the long-time limit, we have
\begin{align}
    P^*_\mathrm{even}(x,t) &\approx
    \left\{
    \begin{array}{lr}
        \frac{x \exp \left(- t \left[ \frac{z^2}{4} + \frac{V_0}{2 k_B T} |z| + \frac{V_0^2}{4 (k_B T)^2} \right] \right) }{\sqrt{\pi t} \left(x - \frac{V_0}{k_B T}t \right)} & \text{if } x < \frac{V_0 t}{k_B T}\\
        - \frac{V_0}{2 k_B T} e^{- \frac{V_0 |x|}{k_B T} } & \text{if } x > \frac{V_0 t}{k_B T}
    \end{array}
\right. \, .
\end{align}
The dynamical phase transition occurs when the integration path envelops the Boltzmann-Gibbs pole. The complete even part of the PDF is then
\begin{align}
    P_\mathrm{even}(x,t) &\approx
    \left\{
    \begin{array}{lr}
    \frac{V_0}{2 k_B T} e^{- \frac{V_0 |x|}{k_B T} } & \text{if } x < \frac{V_0 t}{k_B T}\\
        \frac{x \exp \left(- t \left[ \frac{z^2}{4} + \frac{V_0}{2 k_B T} |z| + \frac{V_0^2}{4 (k_B T)^2} \right] \right) }{\sqrt{\pi t} \left(x - \frac{V_0}{k_B T}t \right)} & \text{if } x > \frac{V_0 t}{k_B T}
    \end{array}
\right.   \, .
\end{align}

    \section{The dynamical phase transition for $0 < \alpha < 1$ \label{eq:phase-transition-general}}

    In our analysis of the $V$-shaped potential, $\alpha=1$, we understood that the negative Boltzmann-Gibbs contribution, necessary to cancel out the equilibrium ground state at large distances, arises from a pole in the complex plane. This is impossible in the odd part of the PDF, which we could easily verify in the $V$-shaped potential (see Eq.\,(\ref{eq:eigenfunction-V-odd})).

    In the case of general exponent $\alpha$, we don't have access to a complete analytical expression for the $k$-density which would allow us to find this pole. As we showed in the previous section, the WKB solution becomes more accurate the longer time $t$ we consider. Despite that, we are still limited to a piece-wise expression, the turning point being the point of switch. The Boltzmann-Gibbs contribution is found when the right WKB solution dominates the integral, and we must perform analytical continuation to find the maximal eigenvalues $\kappa^*$.
    We understand that the pole must be found at $\kappa = 0$ of the analytical continuation of $\mathcal{S}_R(z,\kappa)$. 
    
    Let us express the excited contributions of the even PDF in the usual coordinates. Using that, for the even eigenfunctions, we have $\psi_k^\mathrm{even}(x_0) \approx 1$, the expression for the PDF becomes
    \begin{align}
        P_\mathrm{even}^* (x,t) &= \int_0^\infty \frac{dk}{\pi} \frac{k}{4 A_k^2} \psi_k(x) \psi_k(x_0) e^{-k^2t - \frac{V(x)}{2 k_B T} + \frac{V(0)}{2 k_B T}}  \nonumber \\
        &=\int_0^\infty \frac{dk}{\pi}  e^{-k^2t - \frac{V(x)}{2 k_B T} + \frac{V(0)}{2 k_B T}}
        \left\{
        \begin{array}{lr} 
        \frac{k}{4 A_k} \frac{\exp \left( \int_x^{x_\TP} \sqrt{u_S(y) - k^2} dy \right)}{\left( u_S(x) - k^2 \right)^{1/4}} & \text{if } x < x_\TP(k) \\
        \frac{k}{2 A_k} \frac{\sin \left( \int_{x_\TP}^x \sqrt{k^2 - u_S(y)} dy+ \pi/4\right)}{\left( k^2 - u_S(x)  \right)^{1/4}} & \text{if } x > x_\TP(k)
        \end{array} \right. \nonumber \\
        &= - \frac{2^{3/2}}{Z}  \int_0^\infty \frac{dk}{\pi} 
        \left\{
        \begin{array}{lr} 
        \frac{1}{4\sqrt{k}} \frac{\exp \left( - S_L(x,k) \right)}{\left( u_S(x) - k^2 \right)^{1/4}} & \text{if } x < x_\TP(k) \\
        \frac{1}{2\sqrt{k}} \frac{\mathrm{Im} \exp \left( - S_R(x,k) \right)}{\left( k^2 - u_S(x)  \right)^{1/4}} & \text{if } x > x_\TP(k)
        \end{array} \right. \label{eq:integral-PDF-even}
        \, ,
    \end{align}

    This is the function we are interested in
    \begin{align}
        \int_{k_\TP(x)}^\infty \frac{dk}{\pi} e^{-k^2t - \frac{V(x)}{2 k_B T} + \frac{V(0)}{2 k_B T}} \frac{k}{2 A_k} \frac{\sin \left( \int_{x_\TP}^x \sqrt{k^2 - u_S(y)} dy+ \pi/4\right)}{\left( k^2 - u_S(x)  \right)^{1/4}}
        \, ,
    \end{align}
    which we can write as the imaginary part of
    \begin{align}
        \frac{1}{\pi}  \frac{k}{2 A_k} \frac{e^{-k^2t - \frac{V(x)}{2 k_B T} + \frac{V(0)}{2 k_B T} + i \int_{x_\TP}^x \sqrt{k^2 - u_S(y)} dy+ i \pi/4}}{\left( k^2 - u_S(x)  \right)^{1/4}}
    \end{align}
    as we are hunting for a pole in the region of $k < k_\TP(x)$, we re-write the square roots as
    \begin{align}
        &= \frac{1}{\pi}  \frac{k}{2 A_k} \frac{e^{-k^2t - \frac{V(x)}{2 k_B T} + \frac{V(0)}{2 k_B T} + i \int_{x_\TP}^x i \sqrt{ u_S(y) - k^2} dy+ i \pi/4}}{(-1)^{1/4}\left( u_S(x) - k^2  \right)^{1/4}} \nonumber  \\ \nonumber 
        &= \frac{1}{\pi}  \frac{k}{2 A_k} \frac{e^{-k^2t - \frac{V(x)}{2 k_B T} + \frac{V(0)}{2 k_B T} - \int_{x_\TP}^x \sqrt{ u_S(y) - k^2} dy+ i \pi/4}}{e^{i \pi / 4}\left( u_S(x) - k^2  \right)^{1/4}} \\ \nonumber 
        &= \frac{1}{\pi}  \frac{k}{2 A_k} \frac{e^{-k^2t - \frac{V(x)}{2 k_B T} + \frac{V(0)}{2 k_B T} - \int_{x_\TP}^x \sqrt{ u_S(y) - k^2} dy}}{\left( u_S(x) - k^2  \right)^{1/4}}  \\ 
        &= - \frac{k}{2\pi}  \frac{{2}^{3/2}}{Z \, k^{3/2}} e^{-\sqrt{\frac{\pi}{4}} \left( \frac{\alpha \beta}{2} \right)^{\frac{1}{1-\alpha}} \frac{\Gamma\left( \frac{\alpha}{2 - 2 \alpha} \right)}{\Gamma\left( \frac{1}{2 - 2 \alpha} \right)} \, k^{-\frac{\alpha}{1-\alpha}} - \frac{v(0)}{2}} \frac{e^{-k^2t - \frac{v(x)}{2} + \frac{v(0)}{2} - \int_{x_\TP}^x \sqrt{ u_S(y) - k^2} dy}}{\left( u_S(x) - k^2  \right)^{1/4}} \,  \\
        &= - \frac{1}{\pi}  \frac{{2}^{1/2}}{Z \, k^{1/2}} e^{-\sqrt{\frac{\pi}{4}} \left( \frac{\alpha \beta}{2} \right)^{\frac{1}{1-\alpha}} \frac{\Gamma\left( \frac{\alpha}{2 - 2 \alpha} \right)}{\Gamma\left( \frac{1}{2 - 2 \alpha} \right)} \, k^{-\frac{\alpha}{1-\alpha}}} \frac{e^{-k^2t - \frac{v(x)}{2} - \int_{x_\TP}^x \sqrt{ u_S(y) - k^2} dy}}{\left( u_S(x) - k^2  \right)^{1/4}} \,  .
    \end{align}
    We now return to the function $Q(x)$ in Eq.\,(\ref{eq:WKB-Q-action}). In the exponent, the contribution from the integral is equivalent to $-Q(x)$, while for the left WKB solution we would have $+Q(x)$. This small difference is essential to show that the pole is found at $k=0$. Let us make use of the solution we already found for $Q(x)$ to write
    \begin{align}
        Q(x) \approx  \sqrt{\frac{\pi}{4}} \left( \frac{\alpha \beta}{2 k^\alpha} \right)^{\frac{1}{1-\alpha}} \frac{\Gamma\left( \frac{\alpha}{2 - 2 \alpha} \right)}{\Gamma\left( \frac{1}{2 - 2 \alpha} \right)} - \frac{v(x)}{2} - \frac{1}{4} \ln \left( \frac{k^2}{4 u_S (x)} \right) + O(k^2)\, ,
    \end{align}
    Combining all terms into the exponent, we obtain
    \begin{align}
        &= -\sqrt{\frac{\pi}{4}} \left( \frac{\alpha \beta}{2} \right)^{\frac{1}{1-\alpha}} \frac{\Gamma\left( \frac{\alpha}{2 - 2 \alpha} \right)}{\Gamma\left( \frac{1}{2 - 2 \alpha} \right)} \, k^{-\frac{\alpha}{1-\alpha}} - k^2 t - \frac{v(x)}{2} + Q(x) - \frac{1}{4} \ln \left( u_S(x) - k^2 \right) \nonumber \\
        &= - k^2 t - v(x) - \frac{1}{4} \ln \left( \frac{k^2}{4 u_S (x)} \right) - \frac{1}{4} \ln \left( u_S(x) - k^2 \right) + O(k^2) \, ,
    \end{align}
    now we take the small $k$ limit to expand the second Log in the previous equation
    \begin{align}
        &= - k^2 t - v(x) - \frac{1}{4} \ln \left( \frac{k^2}{4 u_S (x)} \right) - \frac{1}{4} \ln \left( u_S(x)  \right) + O(k^2) \, \nonumber \\
        &= -k^2 t - v(x) - \frac{1}{4} \ln \left( \frac{k^2}{4} \right) + O(k^2) \nonumber \\ 
        &= -k^2 t - v(x) -  \ln \left(\sqrt{ \frac{k}{2}} \right) + O(k^2) \, .
    \end{align}
    We plug the exponent back into the original expression to find
    \begin{align}
        &= - \frac{1}{\pi}  \frac{{2}^{1/2}}{Z \, k^{1/2}} e^{-k^2 t - v(x) -  \ln \left(\sqrt{ \frac{k}{2}} \right) + O(k^2) } \nonumber \\
        &= - \frac{1}{\pi}  \frac{{2}}{Z \, k} e^{-k^2 t - v(x) + O(k^2) } \, .
    \end{align}
    The residue at $k=0$ is
    \begin{align}
       \mathrm{Res} &= - \frac{4 i}{Z} e^{- v(x)} \, ,
    \end{align}
    and therefore, the integral, which is $2 \pi i \, \mathrm{Res}$, is equivalent to $4$ times the Boltzmann-Gibbs equilibrium distribution. The imaginary unit is not important since we are interested in the imaginary part. We also have changed the integration limit from $(k_\TP(x),\infty)$ to $(-\infty,\infty)$. This is done by closing the contour through the imaginary plane, crossing the point of local maxima $\kappa^*$, and will require us to divide the result by a factor of 2. Further, since we are not just enveloping the pole, but crossing it, we must divide by an additional factor of 2. This allows us to recover the Boltzmann-Gibbs equilibrium solution.

    \section{The stretched exponential relaxation \label{sec:moments}}

\begin{figure}
        \centering
        \includegraphics[width=0.55\textwidth]{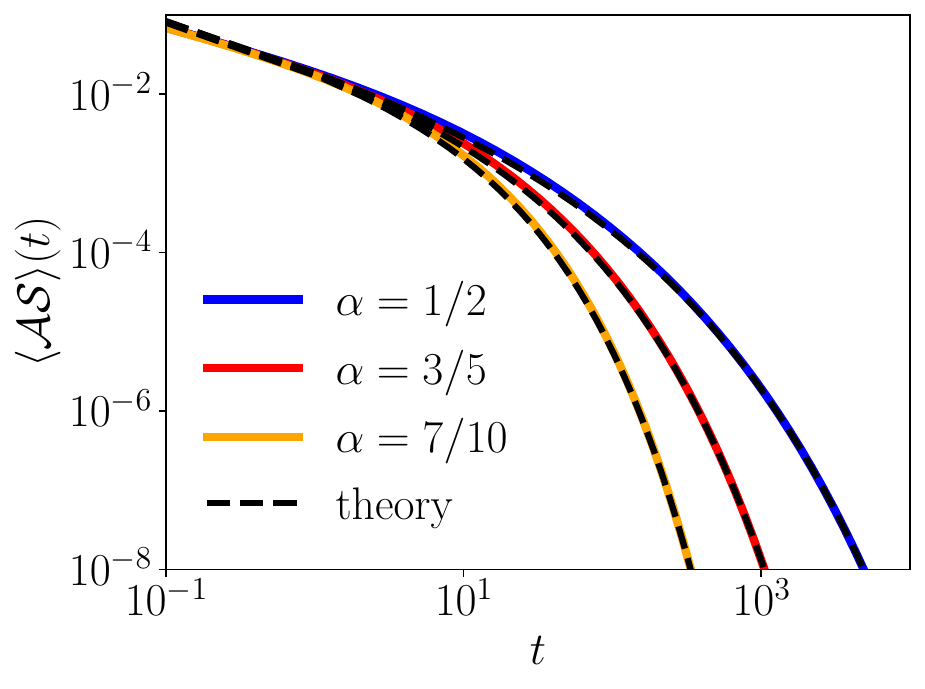}
        \caption{The numerical ensemble average of the asymmetry observable, see Eq.\,(\ref{eq:asymmetry-observable-mean}), as a function of time for three different values of the exponent $\alpha$ (shown in legend). The theoretical prediction in Eq.\,(\ref{eq:asymmetry-observable-mean}) shows perfect agreement for long-times. We have used $\beta=1$ and $x_0=0.04$ initial condition.}
        \label{fig:asymmetry-observable}
    \end{figure}

    The complete eigenfunction expansion allows us to find the complete expression for the PDF. In the long-time limit, the WKB solution for the eigenfunctions becomes more accurate. We also have that the eigenfunctions always decay exponentially, indicating that for calculating observable averages, smaller values of $x$ are more important than larger ones. This means that in the relevant limit $x \ll x_\TP$, we can write the eigenfunctions as
    \begin{align}
        \psi_k(x) & \approx  \frac{x^{1-\alpha}}{\beta \alpha} e^{- \frac{V(0)}{2k_B T}} \exp \left\{ \frac{\beta x^\alpha}{2} - \frac{k^2 x^{2-\alpha}}{\beta \alpha (2-\alpha)} \right\} 
    \end{align}
    For finite values of $x$, the maximum contribution comes from $\kappa^*(z=0)$. We obtain the even part of the PDF as
    \begin{align}
        P_\mathrm{odd} (x,t) & \approx  x_0 \left( \frac{x^{1-\alpha}}{\alpha \beta}  \right) \frac{\nu \I_\mathrm{cont}(0)}{t^{3/2-\nu}} \sqrt{\frac{\gamma - \nu}{4 \pi}} e^{\frac{V(0)}{k_B T}} e^{-\I_\mathrm{cont}(0) t^\nu  - \frac{\I_\mathrm{cont}(0)  x^{2-\alpha}}{\beta (2-\alpha)^2 \, t^{1-\nu}} } \, .
    \end{align}
    See Eq.\,(\ref{eq:critical-kappa-z0}) for the definition of $\I_\mathrm{cont}(0)$. We can now calculate the ensemble average of the asymmetry operator $\mathcal{AS}(x,t)$, defined as
    \begin{align}
        \mathcal{AS}(x) = \left\{
    \begin{array}{rr}
    -1 ~~~ \mathrm{for} ~~~ x > 0 \\
    0 ~~~ \mathrm{for} ~~~ x = 0 \\
    +1 ~~~ \mathrm{for} ~~~ x > 0
    \end{array}
\right. \, ,
    \end{align}
    and the ensemble mean is
    \begin{align}
        \langle \mathcal{AS} \rangle(t) &=  \frac{2 x_0}{\alpha \beta}  \frac{\nu \I_\mathrm{cont}(0)}{t^{3/2-\nu}} \sqrt{\frac{\gamma - \nu}{4 \pi}} e^{\frac{V(0)}{k_B T}} e^{-\I_\mathrm{cont}(0) t^\nu} \int_0^\infty dx \, x^{1-\alpha} e^{  - \frac{\I_\mathrm{cont}(0)  x^{2-\alpha}}{\beta (2-\alpha)^2 \, t^{1-\nu}} } \nonumber \\
        &= \frac{x_0}{\alpha \beta}  \frac{\nu \I_\mathrm{cont}(0)}{t^{3/2-\nu}} \sqrt{\frac{\gamma - \nu}{\pi}} e^{\frac{V(0)}{k_B T}} e^{-\I_\mathrm{cont}(0) t^\nu} \frac{\beta (2-\alpha) t^{1-\nu}}{\I_\mathrm{cont}(0) } \nonumber \\
        &=  \frac{x_0}{t^{1/2}} \sqrt{\frac{\gamma - \nu}{\pi}} \, \, e^{\frac{V(0)}{k_B T}} e^{-\I_\mathrm{cont}(0) t^\nu} \, . \label{eq:asymmetry-observable-mean}
    \end{align}
    The validity of the previous Eq. is shown numerically in Fig.\,\ref{fig:asymmetry-observable}. We extend this result to all odd moments
    \begin{align}
        \langle x^{2n+1} \rangle &=  \frac{x_0}{\alpha \beta}  \frac{\nu \I_\mathrm{cont}(0)}{t^{3/2-\nu}} \sqrt{\frac{\gamma - \nu}{\pi}} e^{v(0)} e^{-\I_\mathrm{cont}(0) t^\nu} \int_0^\infty dx \, x^{2n+2-\alpha} e^{  - \frac{\I_\mathrm{cont}(0)  x^{2-\alpha}}{\beta (2-\alpha)^2 \, t^{1-\nu}} } \nonumber \\
        &= \left[ {x_0} \left( \frac{\beta (2-\alpha)^2}{\I_\mathrm{cont}(0)} \right)^{\frac{2n+1}{2-\alpha}} \sqrt{\frac{\gamma - \nu}{\pi}} e^{v(0)} \Gamma \left( \frac{3+2n-\alpha}{2-\alpha} \right) \right] t^{\frac{8n(1-\alpha) -\alpha^2}{2(2-\alpha)^2}} e^{-\I_\mathrm{cont}(0) t^\nu} \, .
    \end{align}
    The even contributions will converge to the Boltzmann-Gibbs expectation. We can calculate how observables relax to this value using the 
    excited even part of the PDF,
    \begin{align}
        P_\even^*(x,t) & \approx - \frac{1}{Z} \left( \frac{x^{1-\alpha}}{\alpha \beta}  \right) \sqrt{\frac{\gamma - \nu}{\pi t}}  e^{v(0)} e^{-\I_\mathrm{cont}(0) t^\nu  - \frac{\I_\mathrm{cont}(0)  x^{2-\alpha}}{\beta (2-\alpha)^2 \, t^{1-\nu}} } \, ,
    \end{align}
    to obtain the excited even moments
    \begin{align}
        \langle x^{2n} \rangle^* &=  - \frac{2}{Z}  \frac{e^{v(0)}}{\alpha \beta} \sqrt{\frac{\gamma - \nu}{\pi t}}  e^{-\I_\mathrm{cont}(0) t^\nu}  \int_0^\infty dx \, x^{2n+1-\alpha} e^{  - \frac{\I_\mathrm{cont}(0)  x^{2-\alpha}}{\beta (2-\alpha)^2 \, t^{1-\nu}} } \nonumber \\
        &= -\left[  \left( \frac{\beta (2-\alpha)^2}{\I_\mathrm{cont}(0)} \right)^{-1- \frac{2n}{2-\alpha}} \frac{2 e^{v(0)} \Gamma \left( 1 + \frac{2n}{2-\alpha} \right)}{Z(2-\alpha)\alpha \beta} \sqrt{\frac{\gamma - \nu}{\pi}}   \right] t^{\frac{4+3\alpha^2 + 8n - 8 \alpha(n+1)}{2(2-\alpha)^2}} e^{-\I_\mathrm{cont}(0) t^\nu} \, .
    \end{align}
    It is clear that the relaxation of all observables is through a stretched exponential. Remarkably, the rate function, which typically is used to calculate the statistics of rare events, in our case controls the rate of the stretched exponential via $\I_\mathrm{cont}(0) $.

    \section{Path integral approach \label{sec:path-integral}}

    We now present another way of obtaining the scaling rate function. It is possible to express the probability density $P(x,t)$, with initial condition $P(x,0) = \delta(x)$, using a path integral formulation. For a stochastic process that is governed by the Langevin Eq.
    \begin{align}
        \Gamma \dot{x} = - \partial_x V(x) + \eta(t) \, ,
    \end{align}
    where the damping constant $\Gamma$ obeys Einstein's relation, $\Gamma \equiv k_B T / D$, and $\eta(t)$ is an unbiased normal-distributed stochastic force, $\langle \eta(t) \rangle = 0$, with variance
    \begin{align}
        \langle \eta(t) \eta(t') \rangle = 2 k_B T \, \Gamma \, \delta(t-t') \, .
    \end{align}
    The Langevin equation describes the system on the level of the individual possible trajectories. It is possible to extract the PDF $P(x,t)$, of a packet of particles. This PDF is equivalent to one that is obtained from the FPE\,(\ref{eq:FPE}), as these two systems are entirely equivalent.
    
    One alternative way to obtain the probability density $P(x,t)$ is through the path-integral formalism. In this case, the probability density at position $x$ and time $t$ is found by considering the contribution from all possible paths $x(t')$, with initial position $x(0) = 0$, and final position $x(t) = x$.
    \begin{align}
        P(x,t) &= \int \mathcal{D}x(t') e^{- \mathcal{A}[x(t'),\dot{x}(t')]} \label{eq:Pxt-path-integral}
    \end{align}
    where $\mathcal{D}x(t')$ is the Wiener measure, which will weight all possible trajectories and the action $\mathcal{A}[x(t'),\dot{x}(t'),t]$ being
    \begin{align}
        \mathcal{A}\big[ x(t'),\dot{x}(t'),t\big] &= \frac{1}{4} \int_0^t dt' \Big[ \dot{x}(t') - f \big(x(t')\big) \Big]^2 + \frac{1}{2} \int_0^t dt' \, f^{\prime} \big(x(t')\big)  \, , \label{eq:path-integral-action}
    \end{align}
    where we now express the position in units of $\ell$ and time in units of $\ell^2/D$ and we remind that $f(x) \equiv -  \partial_x v(x)$. The second term in the previous equation is due to the Stratonovich interpretation of stochastic calculus. 

    \subsection{Equivalence with the scaling rate function}

    There is a trajectory $x^*(t')$ which minimizes the action in Eq.\,(\ref{eq:path-integral-action}), and therefore is the maximum contribution to the integral in Eq.\,(\ref{eq:Pxt-path-integral}). We can define the action as a function of the final position $x$ and time $t$, as $\mathcal{A}^*(x,t) \equiv \mathcal{A}[x^* (t'),\dot{x}^*(t'),t]$. Our goal here is to show that this action, in the proper scaled variables, is equivalent to the rate function we obtained by solving the differential equation in Eq.\,(\ref{eq:df-rate-function}). 
    
    We are interested in the limit of $t \to \infty$, let us introduce the scaled time as $\tau \equiv t' / t$, so that $\tau \in [0,1]$. Second, we scale the position as $z(\tau) \equiv x(t') / t^\gamma$, and the final position $z \equiv x / t^\gamma$. The action, in these new variables, becomes
    \begin{align}
        \mathcal{A}\big[ x(t'),\dot{x}(t')\big] &= \frac{t^\nu }{4} \int_0^1 d\tau \Big[ \dot{z}(\tau) + \beta \alpha \big( z(\tau) \big)^{\alpha - 1} \Big]^2 = t^\nu \tilde{\mathcal{A}} \big[ z(\tau),\dot{z}(\tau)\big] \, , \label{eq:scaled-action-path-integral}
    \end{align}
    where we identify the scaled action $\tilde{\mathcal{A}} \big[ z(\tau),\dot{z}(\tau)\big]$. To obtain the previous equation, we neglected terms of order $1/t$. We observe the expected scaling for the rate function. In the scaled variables, we have that 
    \begin{align}
        \mathcal{A}^*(x,t)  = t^\nu \tilde{\mathcal{A}}^*(z) \label{eq:scaled-action-optimal-path} \, .
    \end{align} To completely show the equivalence between the action at the maximal trajectory and the rate function, we first construct the equivalent Lagrangian function from Eq.\,(\ref{eq:path-integral-action}),
    \begin{align}
        \mathcal{L}(x,\dot{x}) = \frac{1}{4} \Big[ \dot{x} - f \big(x\big) \Big]^2 + \frac{1}{2} \, f^{\prime} \big(x \big) \, . 
    \end{align}
    The corresponding Hamiltonian function is 
    \begin{align}
        \mathcal{H}(x,p) = p^2 + p \, f(x) - \frac{1}{2} \, f^{\prime} \big(x \big) \, ,
    \end{align}
    where the generalized moment is $p=(\dot x - f(x) ) / 2$. Through the Hamilton-Jacobi equation, we know that the action must respect the equation
    \begin{align}
        - \frac{\partial \mathcal{A}^*}{\partial t} = \mathcal{H} \left(x , \frac{\partial \mathcal{A}^*}{\partial x} \right) = \left( \frac{\partial \mathcal{A}^*}{\partial x} \right)^2 + f(x) \frac{\partial \mathcal{A}^*}{\partial x} - \frac{\partial^2 \mathcal{A}^*}{\partial x^2} \, . \label{eq:hamilton-jacobi}
    \end{align}
    By finally expressing the previous equation in the scaled variables. The partial derivatives become
    \begin{align}
        \frac{\partial \mathcal{A}^*}{\partial t} &= \frac{\partial }{\partial t} \big[ t^\nu \tilde{\mathcal{A}}^*(x/t^\gamma) \big] = \frac{z \gamma \tilde{\mathcal{A}}^{*\prime}(z) - \nu \tilde{\mathcal{A}}^*(z)}{t^{1-\nu}} \, ,
    \end{align}
    and
    \begin{align}
         \frac{\partial \mathcal{A}^*}{\partial x} &=   \frac{\partial}{\partial x}  \left(t^\nu \tilde{\mathcal{A}}^*(x/t^\gamma) \right)  = \frac{1}{t^{\gamma - \nu}} \frac{\partial \tilde{\mathcal{A}}^{*}}{\partial z} \, ,
    \end{align}
    leading to the differential equation
    \begin{align}
        \left[ \frac{\partial \tilde{\mathcal{A}}^{*}}{\partial z}  \right]^2 - \left( \frac{\beta \alpha}{z^{1-\alpha}} + \frac{z}{2-\alpha} \right) \frac{\partial \tilde{\mathcal{A}}^{*}}{\partial z} + \frac{\alpha \tilde{\mathcal{A}}^{*}}{2-\alpha } = 0 \, ,
    \end{align}
    which is identical to the one in Eq.\,(\ref{eq:df-rate-function}) for the rate function. We can finally conclude that
    \begin{align}
        \I(z) = \tilde{\mathcal{A}}^{*}(z) \, .
    \end{align}

    \subsection{Equivalence with the WKB eigenfunctions}

    For a system with fixed energy, that is, $\mathcal{H}(p,x) = E$, we can find the action as
    \begin{align}
        \mathcal{A}^* (x,t) = -  E t + \int_0^x dx' p(x') \, ,
    \end{align}
    which is a solution to the Hamilton-Jacobi equation, see Eq.\,(\ref{eq:hamilton-jacobi}). Alternatively, it can be seen as the minimization of the abbreviated action, the integral term in the previous equation, where the energy is equivalent to the Lagrange multiplier for the time constraint. This can also be clearly seen in the previous equation.

    The two possible values for the generalized moment $p$ can be obtained, as a function of $x$ and $E$, as
    \begin{align}
        p(x) &= - \frac{f(x)}{2} \pm \sqrt{\frac{f(x)^2}{4} + \frac{f'(x)}{2} + E} \nonumber \\
            &= - \frac{f(x)}{2} \pm \sqrt{u_S(x) + E}\, ,
    \end{align}
    where we identified the effective potential of the Schr\"odinger equation inside the square root. The choice of sign in the square root term is set by the initial condition. For a positive value of $x$, we take the positive sign and vice-versa. Performing the change to the scaled variables, the scaled energy, which we define as $\epsilon$, assumes the same scaling as $\kappa^2$, as
    \begin{align}
        E t \to \epsilon t^\nu \, ,
    \end{align}
    and the action becomes
    \begin{align}
        \tilde{\mathcal{A}}^\star (z) &= - \epsilon + \int_0^z dy \, \left( \frac{\alpha \beta}{2}  y^{\alpha-1} + \sqrt{\frac{\beta^2 \alpha^2}{4 y^{2-2\alpha}} + \epsilon(z) } \right)\nonumber \\
        &= - \epsilon + \frac{\beta z^\alpha}{2} + \int_0^z \sqrt{\frac{\beta^2 \alpha^2}{4 y^{2-2\alpha}} + \epsilon(z) } \, \, dy   \, , \label{eq:path-integral-attion-WKB-form}
    \end{align}
    which shows a clear resemblance with Eqs.\,(\ref{eq:SL-scaled}) and (\ref{eq:SR-scaled}), with ${\kappa^\star}^2$ having the same role as $\epsilon$. Next, we will look at the actual optimal paths to clearly show that indeed Eq.\,(\ref{eq:path-integral-attion-WKB-form}) is equivalent to Eqs.\,(\ref{eq:SL-scaled}) and (\ref{eq:SR-scaled}).

    \subsection{Instantons}
    We shall obtain the optimal trajectory, $z^\star(\tau)$, or instanton, that minimizes $\tilde{\mathcal{A}}$. Using the Euler-Lagrange equation, we find that the optimal path must satisfy
    \begin{align}
        \frac{d^2 z^*}{d\tau^2} = - \beta^2 \alpha^2 (1-\alpha) z^{2\alpha - 3}   \, ,
    \end{align}
    with $z^*(0) = 0$ and $z^*(1) = z$. We multiply both sides of the previous equation by $\dot{z}$ to obtain an effective energy description as
    \begin{align}
        \frac{d}{d\tau} \left[ \frac{1}{2} \left( \frac{d {z}^*}{d \tau} \right)^2 - \frac{\beta^2 \alpha^2}{2 {z^*}^{2-2\alpha}} \right] = 0 \longrightarrow \frac{1}{2} \left( \frac{d {z}^*}{d \tau} \right)^2 - \frac{\beta^2 \alpha^2}{2 {z^*}^{2-2\alpha}} = 2 \epsilon(z) \, , \label{eq:energy-balance}
    \end{align}
    where the constant $\epsilon(z)$ is set by the target final position $z$ and plays the same role as the total energy in Eq.\,(\ref{eq:path-integral-attion-WKB-form}). The connection between the action in the previous equation and the action from the WKB approach will be explained in detail further. It is easier to solve this problem backward in time, starting at $z$ until reaching the origin. This is ideal as we are interested in the $z>0$ range, and as $z \to 0$, we have to be more careful to understand the role of the fine structure of the potential.

    From Eq.\,(\ref{eq:path-integral-attion-WKB-form}), we can see that $\epsilon = 0$ is equivalent to the potential, and therefore, the Boltzmann-Gibbs rate function. The actual trajectory is obtained from Eq.\,(\ref{eq:energy-balance}), 
    \begin{align}
        \frac{d z^*}{d \tau} = 2 \sqrt{\frac{\beta^2 \alpha^2}{{4 z^*}^{2-2\alpha}}} = \frac{\beta \alpha}{{z^*}^{1-\alpha}} \, . \label{eq:path-integral-implicity}
    \end{align}
    In the backwards time description, the previous equation is equivalent to a particle that starts at position $z$ with kinetic energy identical to the negative of the potential energy, that is, $\dot{z}^*(1) = f(z)$ (which is a negative value). For $z<z_c$, the particle reaches the origin at a finite time $\delta \tau(z)$, which is obtained directly from Eq.\,(\ref{eq:path-integral-implicity}) as
    \begin{align}
        \delta \tau (z) &= \int_0^z \frac{y^{1-\alpha}}{\beta \alpha} dy = \frac{z^{2-\alpha}}{\beta \alpha (2-\alpha)} = \left[ \frac{z}{z_c} \right]^{2-\alpha} \, .
    \end{align}
    Clearly, from the previous equation, for $z=z_c$, the particle reaches the origin exactly at $\tau=1$. If the particle is closer to the origin than the critical $z_c$, the particle will reach the origin at a finite time and then become stuck at $z=0$, which is also a solution to the Euler-Lagrange equation. 

    The complete solution, for $z<z_c$, can be written as
    \begin{align}
        z^* (\tau) =  \left\{
    \begin{array}{lr}
    0 ~~~ \mathrm{for} ~~~ \tau \leq \tau_0 \\
    z_c \left( \tau - \tau_0 \right)^{1/(2-\alpha)} ~~~ \mathrm{for} ~~~ \tau > \tau_0
    \end{array}
\right.   \, ,
    \end{align}
    where $\tau_0 = \left[ {z}/{z_c} \right]^{2-\alpha}$ is the instant where the particle leaves the origin. In the actual coordinates, being $x$ and $t$, this break at $\tau = \tau_0$ is smooth, as we can see from the numerical integrations of the Euler-Lagrange equations at finite time. The break only occurs in the $t \to \infty$ scaling limit. The action can be calculated directly as
    \begin{align}
        \tilde{\mathcal{A}} (z) &= \frac{1}{4} \int_0^1 d\tau \Big[ \dot{z}^* + \beta \alpha {z^*}^{\alpha - 1} \Big]^2 = \frac{1}{4} \int_{\tau_0}^1 d\tau \Big[ \dot{z}^* + \beta \alpha {z^*}^{\alpha - 1} \Big]^2 \nonumber \\
        &= \frac{z_c^2}{(2-\alpha)^2} \int_{\tau_0}^1 \frac{d \tau}{(\tau - \tau_0)^{\frac{2-2\alpha}{2-\alpha}}} = \beta z^\alpha \, ,
    \end{align}
    which, as expected, is equivalent to the potential. 
    
\begin{figure}
       \centering
       \includegraphics[width=0.7\textwidth]{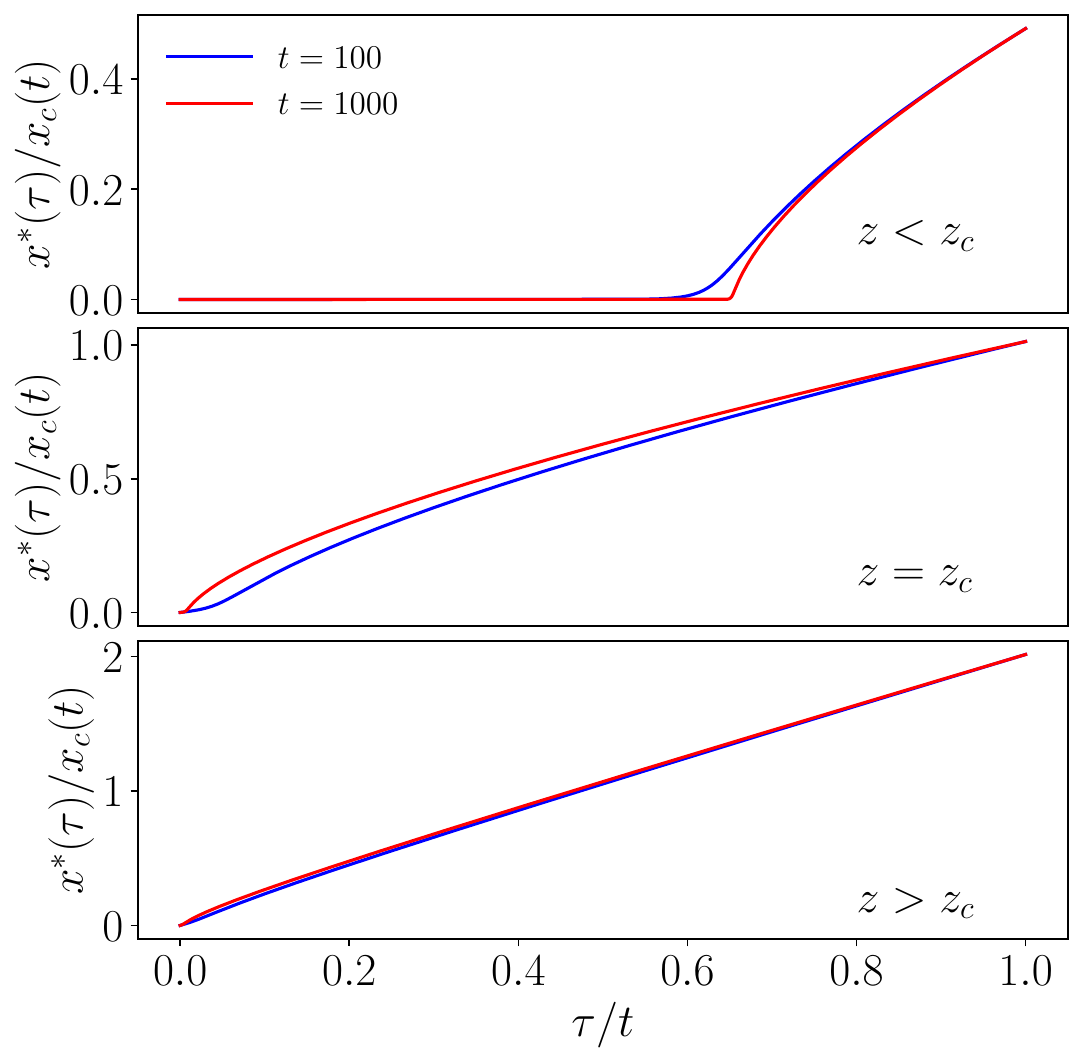}
       \caption{The optimal trajectories obtained numerically for finite times (shown in the top legend, common for all panels). We have used $\alpha=1/2$ and $\beta = 1$.}
       \label{fig:BG-trajectories}
   \end{figure}

    In the other case, for positions larger than the critical point, $z>z_c$, the particle requires the energy to be bigger than zero in order to reach the origin in the available time frame. This sharp transition is the dynamical phase transition: the system goes from the zero energy Boltzmann-Gibbs solution to a finite energy one. And, for very large value of $z$, the necessary energy becomes much greater than the potential and we obtain a straight line solution, which is equivalent to the free-particle. We will go into more detail in the next subsection, but the energy $\epsilon$ is equivalent to $-{\kappa^*}^2$, which is a positive number since for $z>z_c$, the critical point is a purely imaginary number. We show this numerically in Fig.\,\ref{fig:BG-trajectories}.

    \subsection{Instantons for the asymmetric part}

    The results we obtained thus far are consistent with the rate function of the complete solution. However, as we have found in both scaling rate function and eigenfunction approaches, there is also a distinct solution that will govern the odd part of the PDF. It is not immediately clear how to obtain these contributions through the path-integral formulation. Let us start by considering an initial condition that will cause the PDF to contain an odd part. Placing the particles initially at $x=x_0$, that is,
    \begin{align}
        P(x,0) = \delta(x-x_0) = \underbrace{\frac{\delta(x-x_0) + \delta(x+x_0)}{2}}_{\mathrm{Symmetric}} + \underbrace{\frac{\delta(x-x_0) - \delta(x+x_0)}{2}}_{\mathrm{Asymmetric}} \, ,
    \end{align}
    we have the even (symmetric) and odd (asymmetric) contributions. We will now consider the paths for all possible delta functions in the previous equations. When we consider the very long time limit, $t \to \infty$, for positions that scale with $t^\gamma$, the contributions from paths that connect $x_0 \to x$ and $x_0 \to -x$ ($-x_0 \to x$ and $-x_0 \to -x$)  become identical and are equivalent to the instanton trajectories we discussed in the previous subsection, as the respective deltas just add their own contributions. Therefore, the symmetric part will behave as described before, exactly as expected. However, since the asymmetric terms are the difference between deltas, these identical paths will actually be canceled out.

    In that case, we must look for asymmetric paths that never cross the origin, meaning for $+x_0$ ($-x_0$), we will only have optimal paths for $x>0$ ($x<0$). In the scaled variable, $x_0/ t^\gamma \to 0$, and the problem becomes almost identical to the previous one, with the difference that we now want the particle to be as close as possible to $z=0$, without actually touching the origin. For $\epsilon=0$, we saw previously that, if $z<z_c$, the particle has more than enough time to reach the origin. Therefore, there is a smaller (negative) energy value where the particle will reach the region near the origin at exactly the available time.

    Let us calculate the energy for the specific example $z=0$, which we remind that we mean actually near zero. When $z<z_\dagger$, the particle initially has a positive velocity, pointing away from the origin. It then reaches the maximum point, with zero velocity, which as we will see is equivalent to the $z_\TP$, and finally returns to the origin.

\begin{figure}
       \centering
       \includegraphics[width=0.9\textwidth]{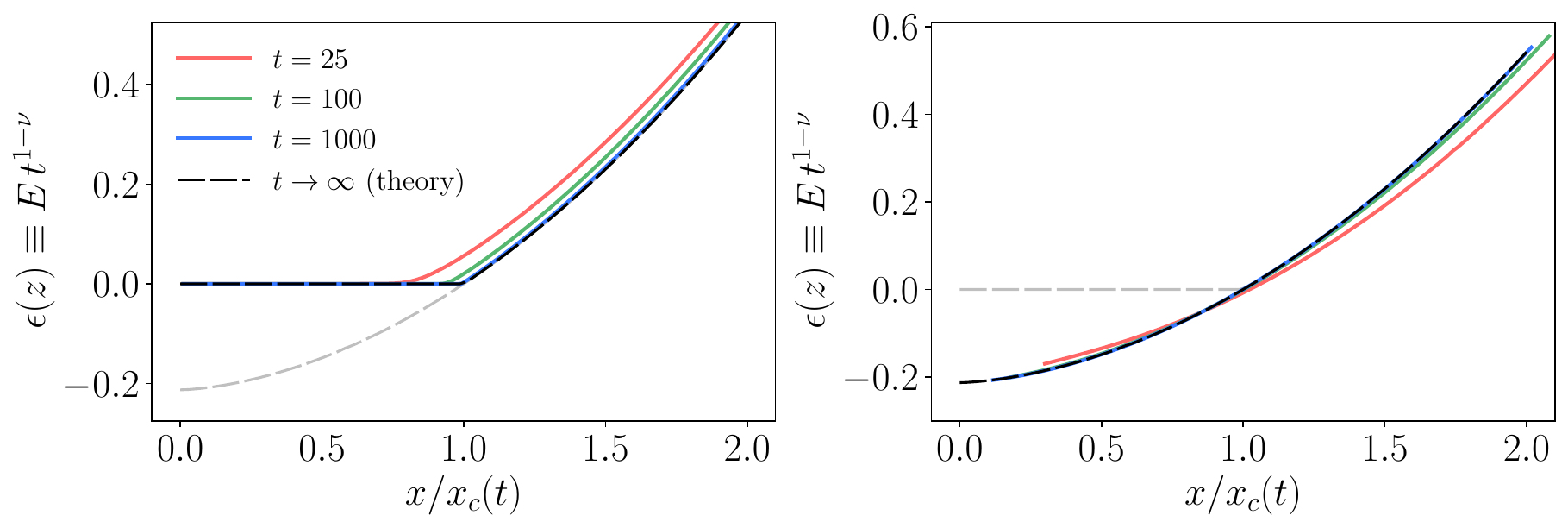}
       \caption{The effective energies for symmetric (left) and asymmetric (right) trajectories for finite times (shown in the caption on the left panel, common for both panels). The dashed lines represent our prediction for infinite time $\epsilon(z)$ for each respective case, we plot both in gray for easy comparison and in black the respective limit. We have used $\alpha=1/2$, $\beta=1$ and, for simplicity, $x_0 = \sqrt{2}$ for the asymmetric case.}
       \label{fig:finite-time-energy}
   \end{figure}

    We use the Eq.\,(\ref{eq:energy-balance}) for the energy balance to write the general expression
    \begin{align}
        \frac{d z^*}{2 \sqrt{\epsilon(z) + \frac{\beta^2 \alpha^2}{{4 z^*}^{2-2\alpha}}}} = d \tau \, . \label{eq:path-integral-implicity1}
    \end{align}
    The integral of $d\tau$ on the right of the previous equation will give us 1, which is the total time. On the left-hand side, the integral must be performed on the ascending path of the trajectory $z \to z_\TP$ and then, with the opposite sign, at the descending side of the trajectory $z_\TP \to 0$. Explicitly, we have
    \begin{align}
        \int_z^{z_\TP} \frac{d z^*}{2 \sqrt{\epsilon(z) + \frac{\beta^2 \alpha^2}{{4 z^*}^{2-2\alpha}}}} + \int_0^{z_\TP} \frac{d z^*}{2 \sqrt{\epsilon(z) + \frac{\beta^2 \alpha^2}{{4 z^*}^{2-2\alpha}}}} = 1 \, ,
    \end{align}
    where we highlight that the value of the energy is related to the turning point as $z_\TP$, as $\epsilon(z) = \beta^2 \alpha^2 / (4 z_\TP^{2-2\alpha})$. In the case of $z=0$,
    \begin{align}
        \int_0^{z_\TP} \frac{d z^*}{\sqrt{- \frac{\beta^2 \alpha^2}{{4 z_\TP}^{2-2\alpha}} + \frac{\beta^2 \alpha^2}{{4 z^*}^{2-2\alpha}}}} = 1 \, ,
    \end{align}
    and we obtain that
    \begin{align}
        z_\TP(0) &= \left[ \left( \frac{2}{\beta \alpha} \right) \sqrt{\frac{\pi}{4}} \left( \frac{\alpha}{1-\alpha} \right) 
\frac{\Gamma \left( \frac{\alpha}{2 - 2\alpha} \right)}{\Gamma \left( \frac{1}{2 - 2\alpha} \right)} \right]^{-1/(2 - \alpha)} \, ,
    \end{align} 
    and the energy is 
    \begin{align}
        \epsilon(0) = - \left[ \frac{\alpha}{1-\alpha} \sqrt{\frac{\pi}{4}} \left( \frac{\beta \alpha}{2} \right)^{1/(1-\alpha)} \frac{\Gamma \left( \frac{\alpha}{2 - 2\alpha} \right)}{\Gamma \left( \frac{1}{2 - 2\alpha} \right)} \right]^{(2-2\alpha)/(2-\alpha)} = - \kappa^*(0)^2 \, ,
    \end{align}
    which is equivalent to Eq.\,(\ref{eq:critical-kappa-z0}) (last equality of the previous equation). Another point we can find exactly is the $z = z_\dagger$, where the particle starts at the top of the trajectory and then we have $z_\TP = z = z_\dagger$. We can find the value of $z_\dagger$ through
    \begin{align}
        \int_0^{z_\dagger} \frac{d z^*}{2 \sqrt{- \frac{\beta^2 \alpha^2}{{4 z_\dagger}^{2-2\alpha}} + \frac{\beta^2 \alpha^2}{{4 z^*}^{2-2\alpha}}}} = 1 \, ,
    \end{align}
    and we immediately recover the $z_\dagger$ obtained from the WKB action, and
    \begin{align}
        \epsilon(z_\dagger) = - \left[ \frac{\alpha}{1-\alpha} \sqrt{\frac{\pi}{16}} \left( \frac{\beta \alpha}{2} \right)^{1/(1-\alpha)} \frac{\Gamma \left( \frac{\alpha}{2 - 2\alpha} \right)}{\Gamma \left( \frac{1}{2 - 2\alpha} \right)} \right]^{(2-2\alpha)/(2-\alpha)} = - \kappa^*(z_\dagger)^2 \, .
    \end{align}
    Using numerical simulations at finite times, we show that, as time increases, the effective energy approaches our predictions both for the symmetric and asymmetric case, see Fig.\,\ref{fig:finite-time-energy}. The dynamical phase transition happens for the symmetric as the effective energy switches from $\epsilon=0$ to a finite value for $z>z_c$.

\begin{figure}
       \centering
       \includegraphics[width=0.7\textwidth]{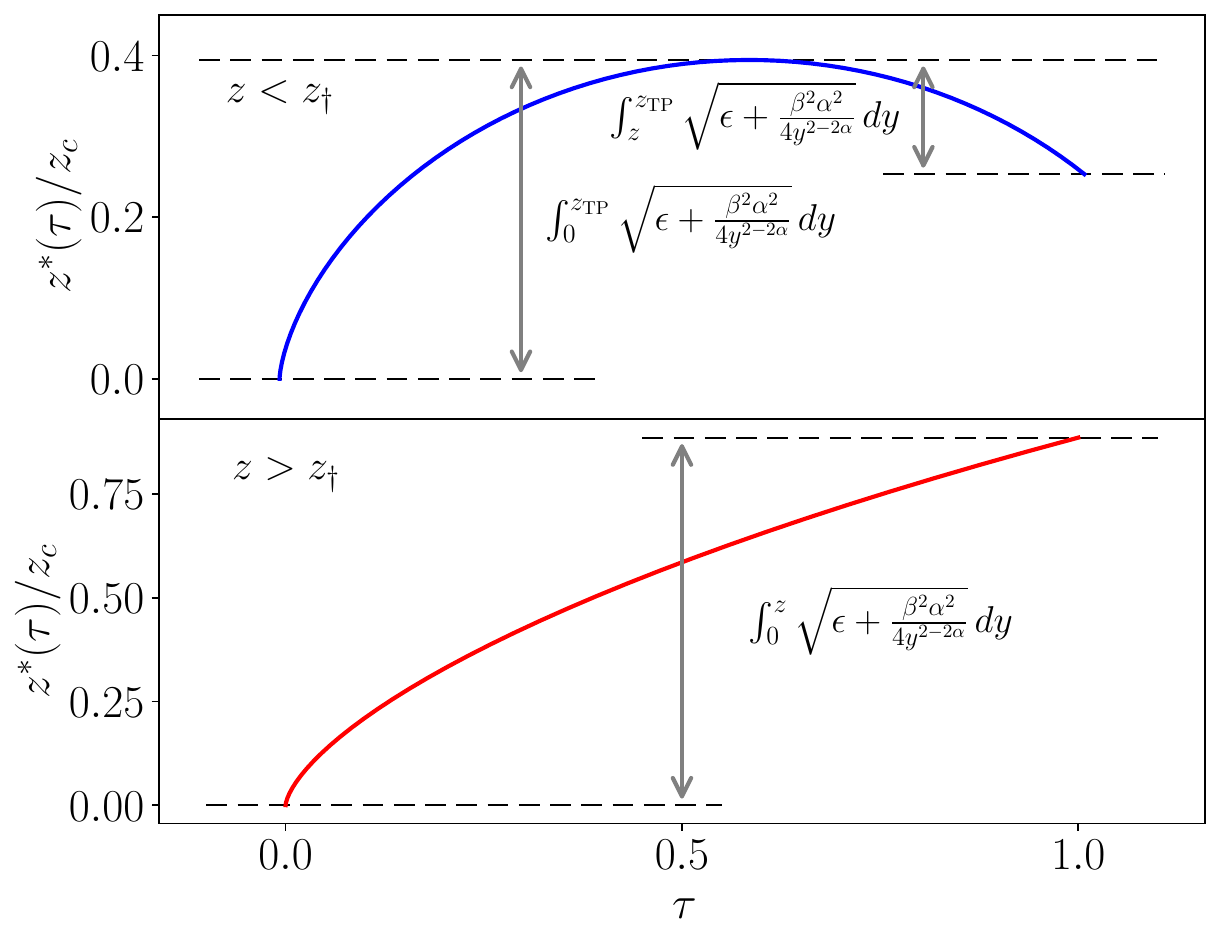}
       \caption{Asymmetric trajectories for $z<z_\dagger$ (upper panel) and $z>z_\dagger$ lower panel. We see clearly that, for $z<z_\dagger$, before reaching the target, the path first goes beyond to $z_\TP > z$, before returning. Each part of this path represents a contribution to the action, shown in the figure. For $z>z_\dagger$, the path goes directly to the target and we only have one contribution to the action. We have used $\alpha=1/2$ and $\beta=1$.}
       \label{fig:excited-trajectories}
   \end{figure}

    From the asymmetric trajectories, we see that there is a shift occurring at $z=z_\dagger$, see Fig.\,\ref{fig:excited-trajectories}. When $z<z_\dagger$, the trajectory reaches a value $z_\TP > z$, before descending into the target $z$. In the eigenfunction approach, this is equivalent to the maximum contribution residing in the left WKB solution. As we see clearly in Eq.\,(\ref{eq:SL-scaled}), the action contains two integral terms, one from 0 to $z_\TP$ and another from $z_\TP$ to $z$. On the other hand, for $z<z_\dagger$, the particle goes directly from the origin to the target $z$. This is also clear in the eigenfunction approach, as we see that in Eq.\,(\ref{eq:SR-z-larger-dagger}), there is only one integral term, from 0 to the target $z$.

    % \cite{naftali_email}
    \section{Final remarks}

In this paper, we present a comprehensive study of weakly bound Brownian particles. We employ three distinct analytical approaches: scaling rate-function solutions, eigenfunction expansions (through the WKB method), and path integral formalism, demonstrating their equivalence. Our findings are summarized by the equation
\begin{align}
\I(z) = \mathcal{S}(z,\kappa^) = \tilde{\mathcal{A}}^(z) , .
\end{align}
The rate function $\I(z)$, obtained using a scaling ansatz, is determined through the non-linear differential equation (\ref{eq:df-rate-function}). The action $\mathcal{S}$ is derived using an eigenfunction approach (see Eqs. (\ref{eq:SL-scaled}) and (\ref{eq:SR-scaled})), which involves a semi-classical approximation (WKB method). We find that this action at the critical point $\kappa^*$ (see Eq.\,(\ref{eq:critical-kappa})) is identical to $\I(z)$. The action $\tilde{\mathcal{A}}^* (z)$ (see Eq.,(\ref{eq:scaled-action-optimal-path})) is obtained through the path-integral method, using the path of minimal action, and is equivalent to the previous two rate functions.

In all three approaches we have a dynamical phase transition occurring at a critical point $z_c$ (Eq.\,(\ref{eq:def-zc})), where the rate function switches between two analytical expressions: the equilibrium rate function $\I_\mathrm{BG}(z)$, and the rate function from the continuum modes $\I_\mathrm{cont}(z)$ (see Fig.\,{\ref{fig:rate-functions}}). This second-order transition is characterized by a discontinuity in the second derivative. We also showed that the appearance of the rate function is connected with parity and initial conditions.

The dynamical exponents $\nu$ and $\gamma$, as functions of $\alpha$ (see Eq.\,(\ref{eq:scaled-exponents})), are consistent across all approaches. These exponents, along with properties such as the dynamical phase transition and stretched exponential relaxation, are expected to remain valid in general dimensions. However, the role of parity in the various rate functions requires careful investigation.

We note that dynamical phase transitions also occur for $1 < \alpha < 2$. In this case, the scaling exponents are larger than one, indicating that $x_c(t)$ grows super-linearly with time, as $\gamma > 1$. For $\alpha \geq 2$, the equilibrium becomes more confining than the Gaussian free-particle, causing the scaling approach to break down and eliminating the dynamical phase transition. The harmonic oscillator represents a special case where the exponents diverge.

Our findings hint that the multi-valued nature of the rate function may be present in more systems. In our analysis using the path integral formalism, it becomes clear that the choice between symmetric and asymmetric paths may lead to different paths and therefore different actions. The rate function at $z=0$, $\I_\mathrm{cont}(0)$, governs the rate of relaxation of the stretched exponential, which is easier to access through experiments. This is remarkable, as the rate function is usually related to fluctuations and rare events. Therefore, a possible extension to this work is to search for physical systems where the rate function can play a meaningful role that can be experimentally probed. 

Finally, given that the Boltzmann measure, $e^{-V(x)/k_B T}$, is sub-exponential, statistical features of the system are described by the big jump principle \cite{Vezzani2019}. This principle may become apparent when time-integrated observables of the underlying processes are considered, although further research is required in this area.

\noindent {\bf Acknowledgements:}
The support of Israel Science Foundation's grant 1614/21 is acknowledged. LD thanks Naftali Smith for insightful conversations.

%\bibliography{bib}


\end{document}